\documentclass{aa}

\usepackage[utf8]{inputenc}
\usepackage{graphicx}
\usepackage{natbib}
\usepackage{hyperref}
\hypersetup{
  colorlinks   = true, %Colours links instead of ugly boxes
  urlcolor     = red, %Colour for external hyperlinks
  linkcolor    = blue, %Colour of internal links
  citecolor   = blue %Colour of citations, could be ``red''
}
\usepackage{array}
\usepackage{makecell}
\usepackage[table,xcdraw]{xcolor}
\usepackage{multirow}
\usepackage{amsmath}

\begin{document}

\title{The role of initial density profiles in simulations of \\coronal wave -- coronal hole interaction}

\titlerunning{Initial density profiles in CW-CH interaction}
\authorrunning{Piantschitsch, I., Terradas, J. et al.}

\author{Piantschitsch, I.$^{1,2,3}$, Terradas, J.$^{1,2}$, Soubrie, E.$^{2,4}$, Heinemann, S.G.$^5$, Hofmeister S.J.$^6$, Soler, R.$^{1,2}$, and Temmer, M.$^3$}

\institute{$^1$Departament de F\'\i sica, Universitat de les Illes Balears (UIB), Carretera de Valldemossa, km 7.5, E-07122 Palma, Spain \\   
$^2$Institute of Applied Computing \& Community Code (IAC$^3$),
UIB, Carretera de Valldemossa, km 7.5, E-07122 Palma, Spain \\
$^3$Institute of Physics, University of Graz, Universit\"atsplatz 5, A-8010 Graz, Austria\\
$^4$Institut d'Astrophysique Spatiale, CNRS, Univ. Paris-Sud, Universit\'e Paris-Saclay, B\^at. 121, 91405 Orsay, France
\\
$^5$Department of Physics, University of Helsinki, P.O. Box 64, 00014, Helsinki, Finland
\\
$^6$Leibniz Institute for Astrophysics Potsdam, An der Sternwarte 16, 14482 Potsdam, Germany
\\
\email{isabell.piantschitsch@uib.es}
}

\date{\today}

\abstract{Interactions between global coronal waves (CWs) and coronal holes (CHs) reveal many interesting features of reflected waves and coronal hole boundaries (CHB) but have fairly been studied so far. Magnetohydrodynamic (MHD) simulations can help us to better understand what is happening during these interaction events, and therefore, to achieve a broader understanding of the parameters involved. In this study, we perform for the first time 2D MHD simulations of a CW-CH interaction including a realistic initial wave density profile that consists of an enhanced as well as a depleted wave part. We vary several initial parameters, such as the initial density amplitudes of the incoming wave, the CH density, and the CHB width, which are all based on actual measurements. We analyse the effects of different incident angles on the interaction features and we use the corresponding time-distance plots to detect specific features of the incoming and the reflected wave. We found that a particular combination of a small CH density, a realistic initial density profile and a sufficiently small incident angle lead to remarkable interaction features, such as a large density amplitude of the reflected wave and a larger phase speed of the reflected wave with respect to the incoming one. The parameter studies in this paper provide a tool to compare time-distance plots based on observational measurements to those created from simulations and therefore enable us to derive interaction parameters from observed CW-CH interaction events that usually cannot be obtained directly. The simulation results in this study are augmented by analytical expressions for the reflection coefficient of the CW-CH interaction which allows us to verify the simulations results in an additional way. This work, with its focus on parameter studies regarding the initial density profile of CWs, is the first of a series of studies aiming to finally reconstruct actual observed CW-CH interaction events by means of MHD-simulations, and therefore, to understand the involved interaction parameters in a comprehensive way.}


\maketitle

\section{Introduction}

Coronal holes (CHs) are regions of low density plasma in the solar corona, which are associated with a primarily open magnetic field configuration leading to the formation of the high-speed component of the solar wind \citep{Cranmer2009, Krieger1973, Nolte1976}. CHs evolve rather slowly, hence, exhibit a stable and durable structure. Being the darkest and least active regions of the Sun, as observed in EUV, their location, their area, their inner structure, and their features along their boundaries are most relevant for understanding the formation of high speed streams and their interaction with the ambient solar wind in the context of  Space Weather \citep{Cranmer2009,riley2015,Hofmeister2018,Heinemann2018,Hofmeister2020,Hofmeister2022,Samara2022}. We can distinguish between CHs that are observed on-disk and CHs that are located off-limb. On-disk CHs can be further subdivided into the group of polar CHs (north and south) which dominate the CH distribution in times of low solar activity, and CHs on the disk center which tend to appear in more active periods of the solar cycle \citep{Cranmer2009}. Both polar and on-disk CHs are relevant regarding the analysis of their interaction with coronal waves (CWs).

Coronal waves (CWs) are defined as large-scale propagating disturbances in the corona and their evolution/propagation can be observed over the entire solar surface. Traditionally, CWs have been denoted as "EIT waves" because they were directly observed for the first time by the Extreme-ultraviolet Imaging Telescope (EIT; \citealt{Delaboudiniere1995}) onboard the Solar and Heliospheric Observatory \citep{Domingo1995}. CWs are often also referred to as EUV waves or coronal bright fronts, and are commonly associated with energetic eruptions such as coronal mass ejections (CMEs) \citep[see e.g.][]{vrsnaklulic2000}. The occurrence of EUV waves is considered to be more strongly related to CMEs than flares. Notwithstanding that marginal link with solar flares, there can be found evidence of an energetic correlation between these three phenomena \citep[see e.g.][]{Wei2014}. For simplicity and due to the fact that other authors that also support a wave model use similar terminology, we hereafter refer to fast-mode MHD waves in the corona as CWs.

The interaction between CWs and CHs results, among other effects, in the formation of reflected, refracted, and transmitted waves (collectively, secondary waves), which confirm their interpretation as fast-mode MHD waves \citep{thompson1998,wang2000,wu2001,Warmuth2004,veronig2010}. Observational evidence for their wave-like behaviour is given by authors who report waves being reflected and refracted at a CH boundary \citep[e.g.][]{Long2008,gopal2009,kienreichetal2013}, transmitted through a CH \citep{liu19,olmedoetal2012} or partially penetrating into a CH \citep[e.g.][]{Veronig2006}. These observational findings are confirmed by numerical simulations describing effects such as deflection, reflection and transmission when a fast-mode MHD wave interacts with a low-density region such as a CH \citep{Piantschitsch2017,afanasyev2018,Piantschitsch2018a, Piantschitsch2018b}. Recent studies on global EUV wave MHD simulations and observational techniques validate the interpretation of global coronal waves as large-amplitude waves \citep{Downs2021}. However, despite these observational and simulation studies, there are several pending questions regarding CW-CH interaction, such as the high and ambiguous phase speed of secondary waves in observational studies of interaction events \citep{gopal2009, podladchikova2019}, or how CW-CH interaction can provide information about CHs themselves, especially about their boundaries and therefore about the physics regarding the interaction between high and slow solar wind streams  \citep{riley2015,Hofmeister2022}.

In order to understand and reconstruct CW-CH interaction events, it is essential to analyse how different initial parameters, such as the initial density amplitude of the incoming wave and its ratio to the amplitude of the depletion as well as the ratio of both of their widths, influence the behaviour of the reflected and transmitted waves. One possible way of reconstructing such an interaction event is the comparison of time-distance plots based on observations with those obtained by magnetohydrodynamic (MHD) - simulations. In several recent observational studies of CWs interacting with CHs time-distance plots are generated in order to analyse some of the kinematic properties of the incoming and the reflected waves \citep{liu19,Chandra2022,Zhou2022,Mancuso2021}. Those initial simulation parameters that best match the time-distance plots of the observations can give us important information about the parameters of the secondary waves in the observed CW-CH interaction event which are usually not available due to the still rather low quality and accuracy of the measurements of CW parameters. This paper, along with its parameter study about initial density profiles in CW-CH interaction, serves as a first step in this reconstruction, that is, we perform simulations about CW-CH interaction events and by doing that, we analyse the influence of critical input parameters, such as the initial density amplitude, on the interaction features. 

What do we know so far from the observations or, in other words, which parameters can be provided by the observations that can be used as input parameters for the simulations? One essential information we need is the density inside of the CH which drops on average $30\%$ to $70\%$ with respect to the quiet Sun \citep{DelZanna1999,DelZanna2018,Saqri2020,Heinemann2021}. Another important parameter for studying CW-CH interactions is the density amplitude of the incoming wave which is usually expressed as a compression factor that has a value of around $1.1$ or less for CWs with moderate speeds waves \citep{Warmuth2015} but can reach compression factors of up to $1.5$ which then are visible even in the lower chromosphere, and are called Moreton waves \citep{Moreton60}. This density amplitude parameter is derived from intensity measurements of coronal waves \citep{Muhr2011} together with the relation between intensity and density that can roughly be described as $\rho/\rho_0 \sim \sqrt{I/I_{0}}$ (see \citet{Warmuth2015}). However, the density profile of the incoming wave is not only characterised by its amplitude but also by the widths of the enhanced pulse and the depleted region at the rear part of the wave which are also considered in the simulation setup. Apart from that, information about the width of the CHB \citep[][]{heinemann2019} is taken into account as well, that is, we differentiate between situations that include a sharp density drop and situations that exhibit a smooth gradient between the CH surrounding and the area inside of the CH. Last but not least, the influence of the incident angle on the features of the secondary waves is part of the parameter study in this paper and allows us to analyse superposition effects within the reflected waves.

In this paper, we study for the first time the influence of a realistic coronal wave density profile, including the enhanced pulse at the wave front as well as the depleted region at its rear part, on CW-CH interaction. In addition to that, we analyse how the CHB width in combination with the CH density and the incident angle affects the properties of secondary waves. The results are analysed especially with regard to the temporal evolution of the density profiles of the reflected waves and the corresponding time-distance plots of the simulated interaction event. These time-distance plots are the first step in understanding and reconstructing observational time-distance plots of interaction events which can be used to derive phase speeds of incoming and reflected waves as well as the density structure close to the CHB before and after the interaction with the CW. The simulation-based time-distance plots created in this paper are supposed to serve as a tool that can be used to understand the observational wave features and therefore understand and also derive parameter values involved in the CW-CH interactions. We specifically point out that this paper, with its focus on parameter studies regarding the initial density profile of CWs, is the first of a series of studies aiming to finally reconstruct actual observed CW-CH interaction events and, hence, to better understand the involved parameters. 

The paper is structured as follows: In Section 2, we list and describe the parameters provided by the observations which are used as initial input parameters for the CW-CH interaction simulations. Section 3 will be dedicated to the numerical setup of the simulations, the description of the algorithm and the equations involved in simulation code. In Section 4, we describe the initial conditions for the 1D case. Section 5 is dedicated to the analysis of the simulation results in the 1D case and we provide analytical expressions for the reflection coefficients in order to compare them to the density amplitudes in the simulations. In Section 6, we present the simulation results for the 2D case, including the role of the initial density profile of the incoming wave, the CH density, the CHB width, and the different incident angles in the interaction process. Moreover, we generate and analyse the corresponding line profiles and simulation-based time-distance plots, and we discuss the analytical expressions and the representative angles for the 2D case. We conclude in Section 7.

\section{Observational parameters}

In the following section we shortly describe the values and the range of the observational parameters that are used as input parameters for the CW-CH interaction simulations. The key parameters for the simulations are the CH density, the density amplitude of the incoming wave, the density profile (enhancement-depletion ratio) of the incoming wave, the CHB width and the incident angle of the incoming wave. 

\begin{figure}[ht!]
\centering\includegraphics[width=0.99\linewidth]{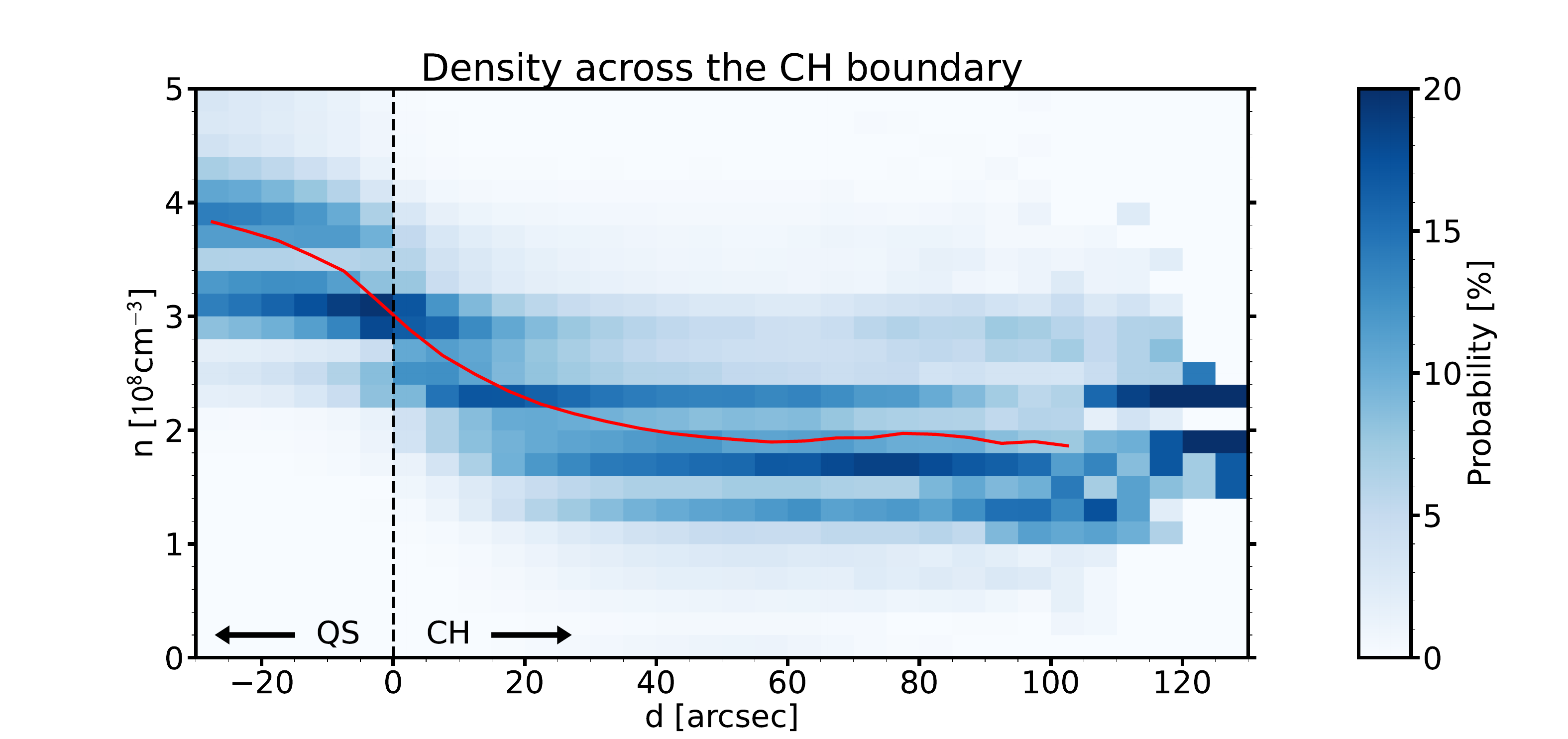}
\caption{Distribution of the electron density as function of distance to the CH boundary [d], adapted from \cite{Heinemann2021}. Each vertical bin is normalized and a darker shade represents a
higher percentage of pixels in the respective density bin. The red line shows the average boundary gradient.}
\label{CHB_trans_area}
\end{figure}

Observational measurements show that the density inside of CHs drops by at least $30\%$ but up to $70\%$ compared to the density values of the quiet Sun \citep[see][]{Doschek1997,DelZanna1999,Saqri2020,Heinemann2021}. Figure~\ref{CHB_trans_area} shows the distribution of electron density values as function of distance to the CH boundary for the dataset presented in \cite{Heinemann2021}. Note that the densities were derived by assuming the same abundances (photospheric abundances) for inside and outside the CH, which is known to not be correct. This means that the density jump near the boundary might be even be significantly larger than shown here. Generally we find that the density drops over a distance of around $20-40"$ to the minimum level, which stays rather constant and homogeneous inside of the CH.

In \citet{Muhr2011} the perturbation of several CWs has been studied and it has been shown that the extracted peak intensity amplitude values of these waves are located in a range between $1.61$ and $1.22$ which are in fact compression factors $n/n_{0}$ of the actual peak intensity. If one neglects the line-of-sight integration effects and the changes in temperature, the observed intensity amplitude can be converted to the density compression factor according to $\rho/\rho_0 \sim \sqrt{I/I_{0}}$ \citep{Warmuth2015}. This implies that the initial density amplitudes of these measurement vary between $1.1$ and $1.2$ which corresponds to the results of \citet{Warmuth2015} where the compression factor of coronal waves with moderate speed is shown to be around $1.1$ which is more representative for a small-amplitudes wave corresponding to a linear or a weakly non-linear wave. Waves exhibiting larger density amplitudes, such as Moreton waves, have much larger compression factors up to $1.5$ and exhbit non-linear behaviour such as the evolution of shocks etc. In this study, we focus on linear and weakly non-linear waves, respectively, which exhibit an initial density amplitude of around $1.1$. 

Besides the density amplitude of the wave front, there are other parameters that define the density profile of the incoming wave. Among them are for instance the width of the pulse that typically increases as the amplitude decrease during its propagation, and also the depleted area at the rear part of the wave (rarefaction region) that usually follows the enhanced pulse. Examples of such density profiles can be seen in \citet{Warmuth2015} and \citet{Muhr2011}. The different amplitudes and widths of the enhanced as well as the depleted part of the incoming wave during its propagation towards the CHB are addressed by varying these parameters in our simulations.

\section{Numerical Setup}

We performed 2.5D simulations of fast-mode MHD waves, which exhibit a realistic initial density profile (enhancement $\&$ depletion), that interact with low density regions representing CHs. The code we used to run the simulations is based on the so-called Total Variation Diminishing Lax-Friedrichs (TVDLF) method, which is a fully explicit scheme and was first described by \citet{toth1996}. We numerically solve the standard MHD equations (see Equations (1)--(3)) and by applying the TVDLF-method, including the Hancock predictor method \citep{VanLeer1984}, we obtain second-order temporal and spatial accuracy. A stable behaviour near discontinuities is guaranteed by applying the so-called Woodward limiter (for details see \citet{VanLeer1979} and \citet{toth1996}). We use transmissive boundary conditions at all four boundaries of the computational box, which is equal to 1.0 both in the $x$- and the $y$-direction. We perform the simulations using a resolution of 300 x 300.

In our simulation code we use the following set of MHD equations including the standard notations for the variables:

\begin{equation}
\frac{\partial{\color{black}\rho}}{\partial t}+\nabla\cdot(\rho \boldsymbol{v})=0
\end{equation}

\textcolor{black}{
\begin{equation}
\frac{\partial(\rho \boldsymbol{v})}{\partial t}+\nabla\cdot\left(\rho \boldsymbol{vv}\right)-\boldsymbol{J}\times \boldsymbol{B}+\nabla p=0
\end{equation}
}

\textcolor{black}{
\begin{equation}
\frac{\partial \boldsymbol{B}}{\partial t}-\nabla\times(\boldsymbol{v}\times \boldsymbol{B})=0
\end{equation}
}

In this study, we consider an idealized case with a homogeneous magnetic field in the $z$-direction and zero pressure all over the computational box, that is, $B_{x}=B_{y}=0$, $p=0$. This is also the reason why no energy equation needs to be included in this setup.

\begin{figure*}
\includegraphics[width=\textwidth]{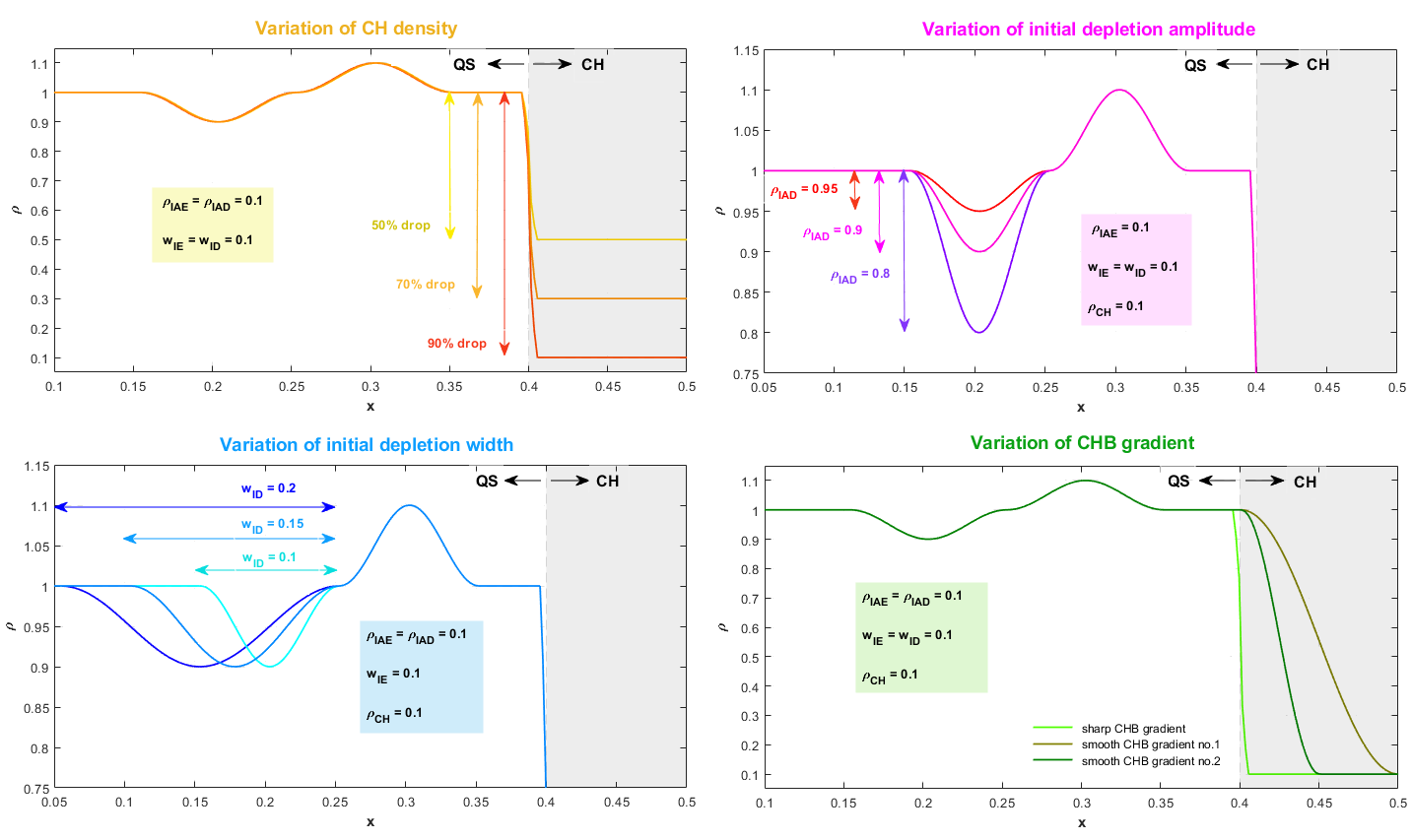}
\caption{Initial conditions for the 1D parameter studies of CW-CH interaction. \textbf{Left upper panel:} Initial density distribution including a realistic density profile for different CH densities ($\rho_{CH}=0.1$, $\rho_{CH}=0.3$, and $\rho_{CH}=0.5$) but with constant values for the initial enhancement and depletion amplitudes ($\rho_{IAE}=1.1$, $\rho_{IAD}=0.9$) and constant initial widths ($w_E=w_D=0.1$). \textbf{Right upper panel:} Initial density distribution for different initial depletion amplitudes ($\rho_{IAD}=0.95,0.9,0.8$) but with a constant initial enhanced amplitude ($\rho_{IAE}=1.1$), constant enhancement and depletion widths ($w_E=w_D=0.1$), and a constant CH density ($\rho_{CH}=0.1$). \textbf{Left lower panel:} Initial density distribution for different initial depletion widths but with constant initial enhancement and depletion amplitudes and a constant CH density. \textbf{Right lower panel:} Initial density distribution for different initial CHB gradients (one sharp gradient and two smooth CHB gradients) and a constant initial wave density profile, that is, constant initial density amplitudes ($\rho_{IAE}=1.1$, $\rho_{IAD}=0.9$) and constant initial widths  ($w_E=w_D=0.1$).}
\label{init_cond_1D_param_var}
\end{figure*}

\section{Initial conditions 1D}

Based on observational measurements, we vary the initial parameters for the simulation setup in the following way:

The first parameter we are going to vary in the course of the simulation process is the CH density. From observational measurements (see Section 2) we know that the density inside of the CH drops by at least $30\%$ but up to $70\%$ compared to the quiet Sun, and the drop might be even significantly larger, considering certain abundance assumptions in the calculation of the density which are known to be erroneous. Hence, as an initial simulation setup we consider the cases of CH densities equal to $0.1$, $0.3$ and $0.5$. In Figure \ref{init_cond_1D_param_var} (top left), one can see the different density drops inside of the CH, whereas the other parameters are kept constant, that is, the amplitude of the density enhancement is equal to $1.1$, the amplitude of the depletion is equal to $0.9$, the widths of the enhancement and the depletion are both equal to $0.1$, and the CHB is considered to be a step function (sharp gradient). 

The enhanced part of the initial wave is excited in the following way for the case of variation of CH density:
\begin{equation}
     \rho(x) = 
    \begin{cases}
        \Delta\rho\cdot \cos^2(\pi\frac{x-x_0}{\Delta x})+\rho_0 & 0.25\leq x\leq0.35 \\
        \qquad 0.1  \lor 0.3 \lor 0.5  & \:\:\:\quad x\geq0.4 \\
        \qquad \qquad1.0 & \:\:\qquad\text{else}
    \end{cases}
    \label{def_rho}
\end{equation}

where $\triangle\rho= 0.1$, $x_0=0.3$, $\triangle x=0.1$, and $\rho_0=1.0$.

\begin{equation}
    v_x(x) = 
    \begin{cases}
        2\cdot \sqrt{\frac{\rho(x)}{\rho_0}} -2.0& \quad0.25\leq x\leq0.35 \\
        \:\:\qquad \qquad0 & \:\:\qquad\quad\text{else}
    \end{cases}
    \label{def_v}
\end{equation}

\begin{equation}
    B_z(x) = 
    \begin{cases}
        \:\:\rho(x) & \quad0.25\leq x\leq0.35 \\
        \:\: 1.0 & \:\:\qquad\quad\text{else}
    \end{cases}
    \label{def_B}
\end{equation}

\begin{equation}
B_{x}=B_{y}=0,\qquad\quad0\leq x\leq0.5
\end{equation}

\begin{equation}
v_{y}=v_{z}=0,\qquad\quad0\leq x\leq0.5.
\end{equation}

The rear and depleted part of the incoming wave is excited in an analogous way, simply by choosing $\triangle\rho= -0.1$ and $x_0=0.2$. The range in which the density, the velocity and the magnetic field values are defined is shifted to $0.15\leq x\leq0.25$.

The second parameter within the simulation setup that will be varied is the initial depletion amplitude. Figure \ref{init_cond_1D_param_var} (top right) shows the values of the depletion amplitude that range from $0.8$ through $0.9$ to $0.95$, whereas the background density is equal to 1.0 and the other parameters are kept constant, that is, the CH density is equal to $0.1$ ($\rho_{CH}=0.1$), the width of the depletion and the width of the enhancement are both equal to $0.1$ ($w_{E}=w_{D}=0.1$), the initial enhancement amplitude is equal to $1.1$ ($\rho_{IAE}=1.1$), and the CHB is again considered to be a step function (sharp gradient). These values correspond to $\triangle\rho= 0.1$  for $0.25\leq x\leq0.35$, $\triangle\rho= -0.2 \lor -0.15 \lor -0.1$ for $0.15\leq x\leq0.25$ , $\rho_0=1.0$, $\rho(x)=0.1$ for $x\geq 0.4$, and $\triangle x=0.1$ in Equations (\ref{def_rho}) - (\ref{def_B}).

The third parameter that needs to be varied is the initial depletion width. In Figure \ref{init_cond_1D_param_var} (bottom left) one can see that the width ranges from $0.1$ through $0.15$ to $0.2$ (corresponds to $\triangle x= 0.1 \lor 0.15 \lor 0.2$ and $x_0= 0.2 \lor 0.175 \lor 0.15$ in Equation (\ref{def_rho}) for the depleted part of the incoming wave), whereas the amplitude for the enhancement and the depletion are kept constant ($\rho_{IAE}=1.1$ and $\rho_{IAD}=0.9$). The width of the initial enhancement is constant as well ($w_{E}=0.1$) and the CH density is equal to $\rho_{CH}=0.1$. Again, as in the setups for the variation of CH density and depletion amplitude, the CHB exhibits a sharp gradient structure.

The fourth and last parameter that gets varied in the course of this parameter study is the CHB width, that is, the area within the density drops from the QS value down to a certain minimum value inside of the CH. In Figure \ref{init_cond_1D_param_var} (bottom right) one can see that in our simulation setup we use three different options for the CHB gradient, first, a sharp gradient, like it is used in the other parameter variations, second, a gradient that exhibits a smoother transition from the QS to the interior of the CH than the sharp gradient (smooth CHB gradient no.1), and third, a very smooth transition (smooth gradient no. 2). The rest of the parameters, all of which describe the density profile of the incoming wave, remain constant in the initial setup, that is, $\rho_{IAE}=1.1$, $\rho_{IAD}=0.9$, $w_E=w_D=0.1$, and $\rho_{CH}=0.1$.

\section{Simulation results 1D}

\subsection{Purely enhanced pulse versus realistic density profile}

As we know from observations, the density profile of a CW that moves towards a CH does not only consist of a purely enhanced pulse but also of a depletion that propagates behind the enhanced part (rarefaction region; e.g. \citet{Muhr2011}). However, in numerical studies about CW-CH interaction, so far only waves exhibiting a purely enhanced pulse were considered which do not reflect the actual wave profiles from the observations though. In this section, we show how significantly the results for the reflected waves change when a more realistic density profile, including an enhanced as well as a depleted part, is used to represent the incoming wave. In addition to that, we analyse the simulation results for different initial density profile setups, CH densities and different CHB gradients.

\begin{figure*}[ht!]
\includegraphics[width=0.99\textwidth]{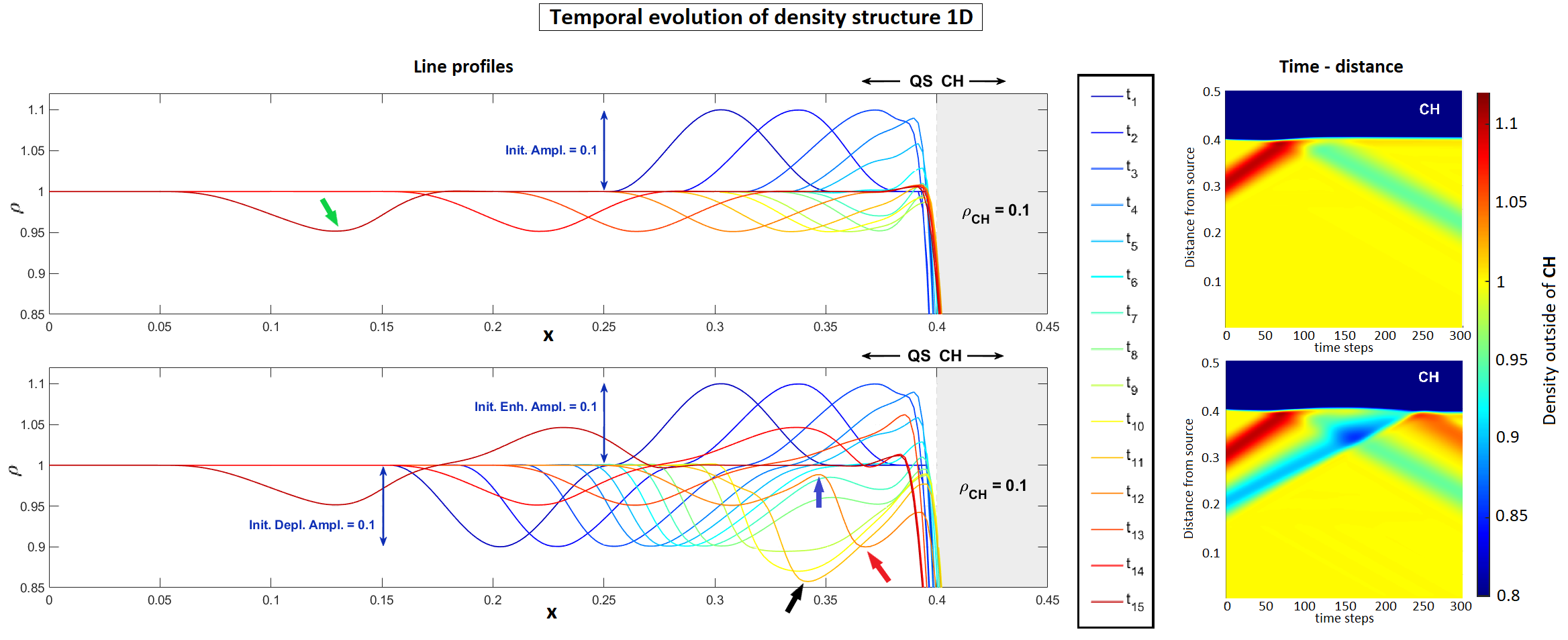}
\caption{Temporal evolution of the density distribution of a CW-CH interaction for the 1D case. The upper panel shows the case of an incoming wave that consists of a purely enhanced wave pulse, whereas the lower panel depicts the situation of a more realistic initial density profile, including an enhanced as well as a depleted wave part. Next to the line profiles one can see the corresponding time-distance plots. The black and the green arrows point at the minimum density values of the simulated interaction which show good agreement with the analytical values based on the reflection coefficient in Equation (\ref{srinc}). The blue and the red arrows show where the reflected wave, that still exhibits negative density values at that time, starts getting pushed towards the background density level.(The temporal evolution of both situations can also be seen in Movie 1 which is attached to this paper.)}
\label{compare_1D_single_pulse_realist_profile}
\end{figure*}

Figure \ref{compare_1D_single_pulse_realist_profile} shows the temporal evolution of the density distribution of a CW-CH interaction including first, a purely enhanced wave pulse (upper panel), and second, including a realistic density profile, that is, the incoming wave consists of an enhanced as well as a depleted part (lower panel). Next to the line profiles in Figure \ref{compare_1D_single_pulse_realist_profile} one can find the corresponding time-distance plots (at the right side). These time-distance plots (and all the following time-distance plots in this paper) show the density profile of the incoming wave (along the $y$-direction and measured as distance from CW source) as a function of the time ($x$-direction) and allow therefore a direct comparison between the properties of the incoming wave and those of the reflected one.

The temporal evolution of the purely enhanced incoming pulse in Figure \ref{compare_1D_single_pulse_realist_profile} results in a reflected wave with density values below the background density (negative amplitude; see $t_{7}$ - $t_{15}$). This is in accordance with what we know about the change of phase in the 1D case (see also \citet{Piantschitsch2020}). The reason for the smaller absolute value of the reflected wave is the fact that a certain part of the incoming wave enters the CH, and hence, there is no total reflection. The phase changing behaviour can be a different one for 2D cases due to superposition effects of the reflected waves, though. As theoretical results in \citep{Piantschitsch2021} show, depending on the CH density and the incident angle, the reflected wave can exhibit an enhanced or a depleted density amplitude, however, for now we stick to the 1D case in which a clear phase changing behaviour can be observed. The corresponding time-distance plot in Figure \ref{compare_1D_single_pulse_realist_profile} exhibits this behaviour too. However, observational time-distance plots do not show such simple structures as in the time-distance plot in the upper panel of Figure \ref{compare_1D_single_pulse_realist_profile}, especially not shortly before and after the actual interaction with the CH. 

Therefore, we analyse the temporal evolution of the more realistic initial density profile, including a depletion at its rear part, which shows a quite different behaviour compared to a purely enhanced pulse. Until the front (enhanced) part of the incoming wave starts interacting with the CH, the shape of the profile stays more or less constant (no strong shock-evolution due to linear and weakly-nonlinear incoming waves). Starting with the actual interaction, the temporal evolution of the density structure gets more complex than in the case of a purely enhanced incoming wave. The incoming enhanced wave part gets reflected and tries to undergo a phase change as in the purely enhanced case and interacts on the way into the negative $x$-direction with the depleted part of the still incoming wave which eventually leads to a large negative amplitude in the reflected wave part (see black arrow in $t_{11}$ in the lower panel of Figure \ref{compare_1D_single_pulse_realist_profile}). In the next step, when the depleted part of the incoming wave starts interacting with the CH and, hence, starts entering the phase changing process, the depleted part of the reflected wave gets larger again (see red arrow in $t_{12}$) and one part of the reflected wave starts getting pushed towards the background density level (see blue arrow in $t_{12}$) until it reaches amplitude values above background density level (see $t_{13}$). Eventually, the reflected wave, consisting of a depleted (moving ahead) and an enhanced part (moving behind) travels into the negative $x$-direction with smaller amplitudes than the incoming wave (see $t_{14}$ and $t_{15}$). The resulting reflected wave shows a clear structure, similar to the incoming wave, consisting of an enhanced as well as a depleted part, however, the interaction between the reflected and still incoming wave parts is a more complex one and crucial to understand the features in observational time-distance plots. By comparing the two time-distance plots in Figure \ref{compare_1D_single_pulse_realist_profile}, one can see the obvious difference between situations including a purely enhanced pulse and situations in which an enhanced as well as a depleted wave part is used to represent the incoming wave. It is important to keep this difference in mind when it comes to study observational time-distance plots of CW-CH interaction events in further studies.  

\subsection{Analytical expressions for the reflection coefficient}

The simulation results shown in Figure \ref{compare_1D_single_pulse_realist_profile} are consistent with the analytical expressions for the so-called reflection coefficient that have been derived in \citet{Piantschitsch2021}. The reflection coefficient, $R$, describes the changes of the amplitude of the reflected wave with respect to the incoming wave.

\begin{equation}\label{srinc}
    R(\theta_{\rm I},\rho_{\rm c})=\frac{\rho_{\rm c} \sin\theta_{\rm I} - f(\theta_{\rm I},\rho_{\rm c})}
    {
    \rho_{\rm c} \sin\theta_{\rm I} +f(\theta_{\rm I},\rho_{\rm c})},
\end{equation}

where $\rho_{\rm c}$ is the density contrast (which is the ratio of the density inside and outside of the CH), $\theta_{\rm I}$ is the incident angle and

\begin{equation}\label{ffunct}
    f(\theta_{\rm I},\rho_{\rm c}) =
    \begin{cases}
    \sqrt{\rho_{\rm c}-\cos^2\theta_{\rm I}} & {\rm if} \quad \theta_{\rm I}\ge  \cos^{-1} \left(\sqrt{\rho_{\rm c}}\right)   \\
    -i\,\sqrt{\cos^2\theta_{\rm I}-\rho_{\rm c}}       & {\rm otherwise.}
    \end{cases}
\end{equation}

We can see that the reflection coefficient depends on only two parameters, the density contrast, $\rho_{\rm c}$, and the incident angle, $\theta_{\rm I}$. Another result from \citet{Piantschitsch2021} is the fact the density perturbation ratio can be written as  

\begin{eqnarray}\label{rho_refl_ratio}
    \frac{\rho^{-}_1}{\rho^{+}_1}=R,
\end{eqnarray}

where $\rho^{-}_1$ denotes the density amplitude of the reflected wave (negative $x$-direction), and $\rho^{+}_1$ denotes the density amplitude of the incoming wave (positive $x$-direction), both propagating in the first medium, that is, the quiet Sun in our case. These equations provide information about how much the density amplitude of the reflected wave decreases with respect to the incoming one.

The analytical results can, in a first step, be compared to the simulation results shown in Figure \ref{compare_1D_single_pulse_realist_profile}. If we assume an incident angle of $90^\circ$ ($\theta_{\rm_I}=90^\circ$), and a density contrast of $\rho_{\rm c}=0.1$ ($\rho_{\rm c}=\rho_{CH}=0.1$ in this case because the density outside of the CH is assumed to be $1.0$), we obtain a reflection coefficient of $R\approx-0.5$ from the analytical expressions. This value is consistent with the 1D simulation of the purely enhanced incoming wave (upper panel of Figure \ref{compare_1D_single_pulse_realist_profile}) where one can see that the final reflected (and smallest) density amplitude is approximately half of the incoming density amplitude (see green arrow). In the case of a more realistic initial density profile (lower panel in Figure \ref{compare_1D_single_pulse_realist_profile}) the simulations are also consistent with the analytically derived reflection coefficient. In this case, one can see that when the reflected wave (reflected depletion) interacts with the still incoming depletion, the sum of the incoming depleted density amplitude $\rho_{IAD}$ and the reflection coefficient $R$ times the incoming enhanced density amplitude $\rho_{IAE}$ is approximately $0.85$ (that is, $\rho_{IAD} + R(\rho_{IAE}-\rho_0)\approx0.85$, see black arrow), the density value that can be seen in the lower panel of Figure \ref{compare_1D_single_pulse_realist_profile}. These results show clearly the consistency between the 1D simulation results and the analytical expressions for the reflection coefficient. 

\begin{table}[ht]
\centering
\resizebox{0.45\textwidth}{!}  
{
\begin{tabular}[t]{c|c|c|}
\cline{2-3}
\multicolumn{1}{l|}{}                                                                                & \begin{tabular}[c]{@{}c@{}}Purely enhanced\\  incoming wave\end{tabular} & \begin{tabular}[c]{@{}c@{}}Realistic initial\\ density profile\end{tabular} \\ \hline
\multicolumn{1}{|c|}{\begin{tabular}[c]{@{}c@{}}Initial incoming \\ density amplitude\end{tabular}}  & $\rho_{IA} = 1.1$                                                            & \begin{tabular}[c]{@{}c@{}}$\rho_{IA} = 1.1$\\ $\rho_{ID} = 0.9$\end{tabular}       \\ \hline
\multicolumn{1}{|c|}{\begin{tabular}[c]{@{}c@{}}Minimum interaction \\ density value\end{tabular}}   & $0.95$                                                                     & $0.85$                                                                        \\ \hline
\multicolumn{1}{|c|}{\begin{tabular}[c]{@{}c@{}}Final depletion \\value of reflection\end{tabular}} & $0.95$                                                                     & $0.95$               \\ \hline
\end{tabular}
}

\vspace{10px}
\caption{Initial, minimum, a final density amplitude values for the a purely enhanced incoming wave and a realistic initial density profile.}

\label{table}
\end{table}

In Table \ref{table} one can see the initial, the minimum, and the final depleted density amplitude values for the purely enhanced incoming wave and for the realistic initial density profile in the 1D case. It can be seen that the main difference between the two situations is the minimum interaction density value. In the situation of the purely enhanced incoming wave, the density value is never below $0.95$ which is also the final depleted value of the reflection. The interaction including a realistic initial density profile, on the other hand, leads to a minimum density value of $0.85$ which is the reason why we see the depleted (dark blue) area in the center of the second time-distance plot in Figure \ref{compare_1D_single_pulse_realist_profile}. However, the final depleted density value in this case is equal to $0.95$ and, therefore, has the same value as found in the interaction with the purely enhanced incoming wave.

\subsection{Parameter studies in 1D}

\begin{figure*}
\includegraphics[width=\textwidth]{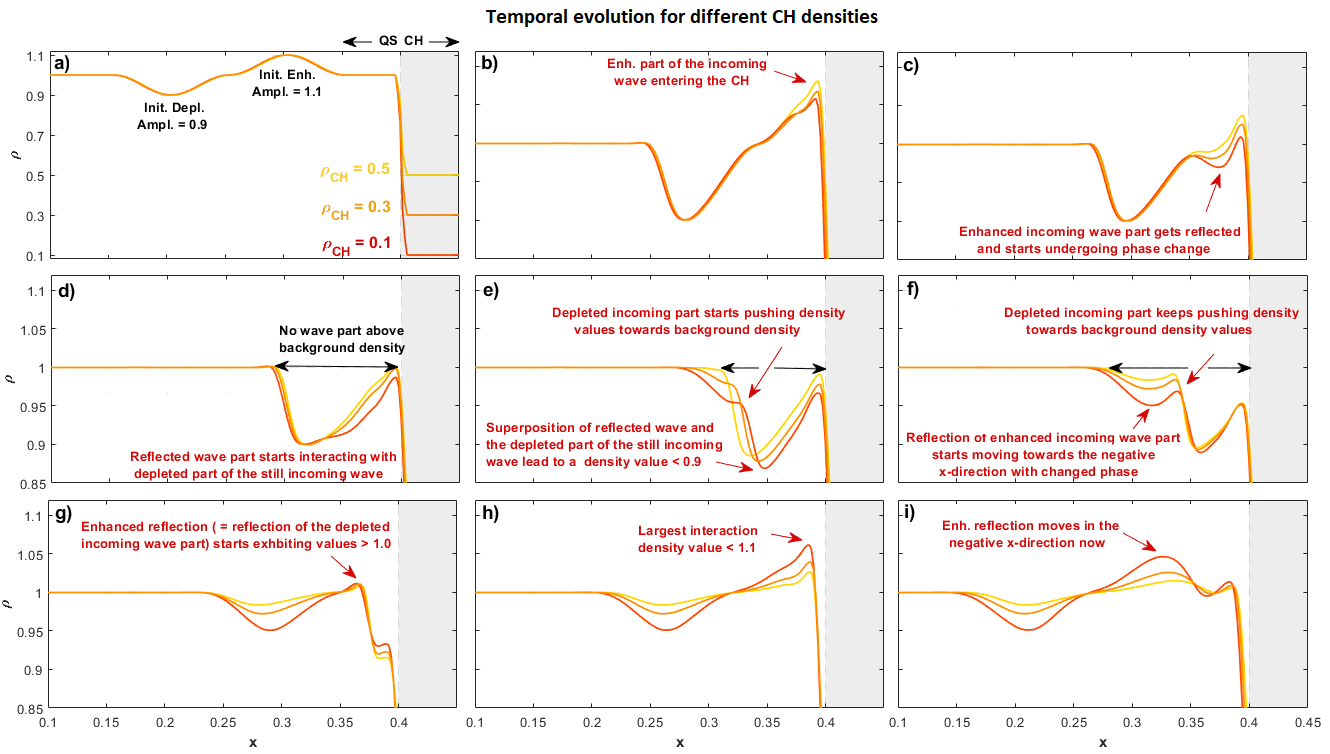}
\caption{Temporal evolution of the density distribution of a CW-CH interaction including a realistic density profile for different CH densities ($\rho_{CH}=0.1$, $\rho_{CH}=0.3$, and $\rho_{CH}=0.5$) but with constant values for the initial enhancement and depletion amplitudes ($\rho_{IAE}=1.1$, $\rho_{IAD}=0.9$) and constant initial widths ($w_E=w_D=0.1$). }
\label{1D_var_CH_density}
\end{figure*}

\begin{figure*}
\includegraphics[width=\textwidth]{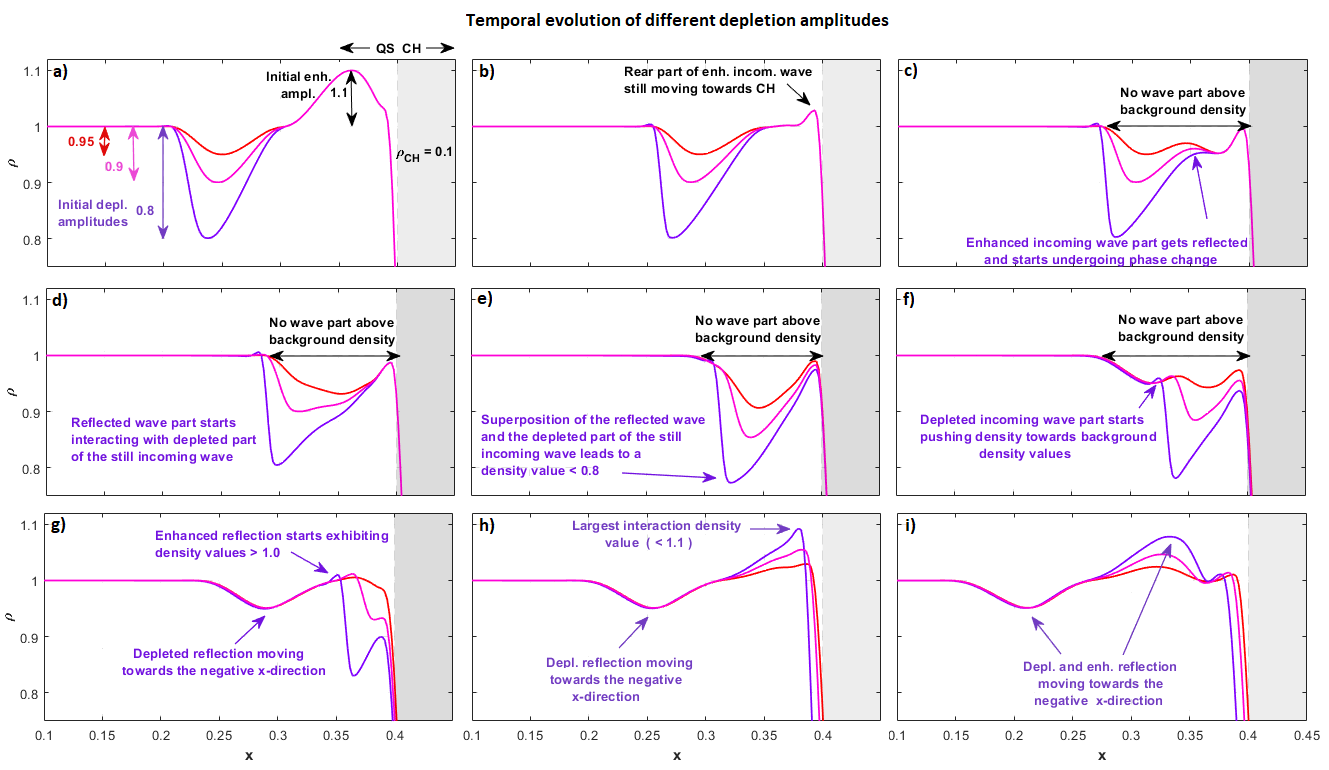}
\caption{Temporal evolution of the density distribution for different initial depletion amplitudes ($\rho_{IAD}=0.95,0.9,0.8$) but with a constant initial enhanced amplitude ($\rho_{IAE}=1.1$), constant enhancement and depletion widths ($w_E=w_D=0.1$), and a constant CH density ($\rho_{CH}=0.1$).}
\label{1D_var_depl_ampl}
\end{figure*}

\begin{figure*}
\includegraphics[width=\textwidth]{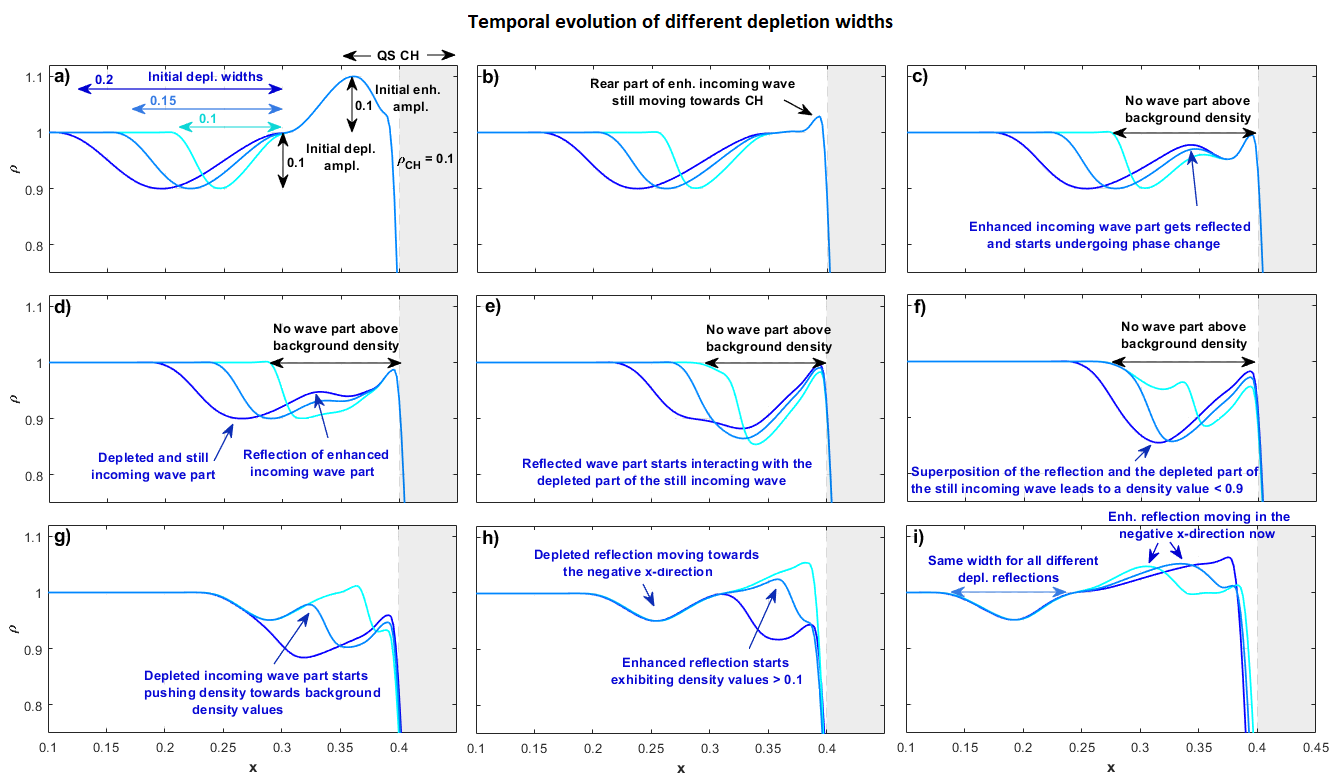}
\caption{Temporal evolution of the density distribution for different initial depletion widths but with constant initial enhancement and depletion amplitudes and a constant CH density. }
\label{1D_var_depl_widths}
\end{figure*}

\begin{figure*}
\includegraphics[width=\textwidth]{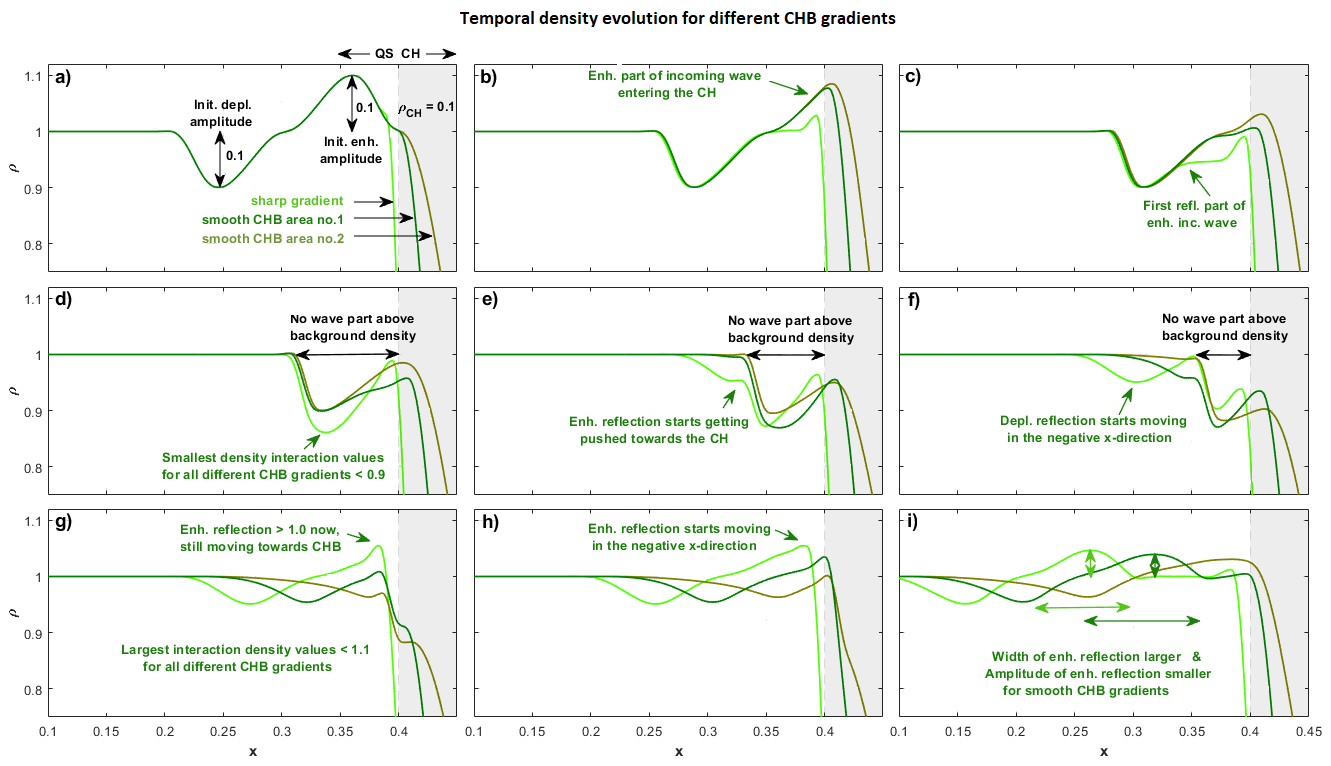}
\caption{Temporal evolution of the density distribution for different initial CHB gradients (one sharp gradient and two smooth CHB gradients) and a constant initial wave density profile, that is, constant initial density amplitudes ($\rho_{IAE}=1.1$, $\rho_{IAD}=0.9$) and constant initial widths  ($w_E=w_D=0.1$).}
\label{1D_var_CHB_gradient}
\end{figure*}

In this subsection, we vary the different initial parameters in the 1D simulation (CH density, depletion amplitude, depletion width, and CHB width) and we analyse the temporal evolution of the density profiles and the corresponding time-distance plots.

In Figure \ref{1D_var_CH_density} one can see the temporal evolution of a CW-CH interaction including a realistic density profile for different CH densities. Figure \ref{1D_var_CH_density}a depicts the initial setup for this interaction process including three different cases for the CH density ($\rho_{CH}=0.1$, $\rho_{CH}=0.3$, and $\rho_{CH}=0.5$). The initial enhanced amplitude, $\rho_{IAE}$, is equal to $1.1$, the initial depleted amplitude, $\rho_{IAD}$, is equal to 0.9, and the widths of enhanced and depleted incoming wave, $w_E$ and $w_D$, are both equal to $0.1$. In Figure \ref{1D_var_CH_density}b one can see how one part of enhanced incoming wave starts entering the CH while another part starts getting reflected. Figure \ref{1D_var_CH_density}c shows how the reflected part of the enhanced incoming wave changes its phase and exhibits density values below background density. The subsequent superposition of the reflected wave (with changed phase) and the depleted part of the still incoming wave leads to density values below $0.9$, which are smaller than the initial depletion amplitude of the incoming wave (see Figure \ref{1D_var_CH_density} d) and e)). When the rear (and depleted) part of the incoming wave starts interacting with the CH, it starts pushing some part of the wave first towards background density (see e) and f)) and eventually above background density ($>1.0$, see g)). The final reflected density profile consists again of a depleted (moving ahead) and an enhanced part (moving behind) but with smaller amplitudes than the ones of the incoming wave. Small CH densities lead to large amplitudes in the reflected wave (see Figure \ref{1D_var_CH_density}i) and also to more visible interaction features during the superposition of the reflected and the still incoming wave (see e) and f)).

Figure \ref{1D_var_depl_ampl} shows the temporal evolution of a CW-CH interaction including a constant enhanced initial amplitude ($\rho_{IAE}=1.1$), a constant CH density ($\rho_{CH}=0.1$), constant initial widths ($w_E=w_D=0.1$) but different initial depletion amplitudes ($\rho_{IAD}=0.95$, $\rho_{IAD}=0.9$, and $\rho_{IAD}=0.8$). The interaction process in this case is similar to the one with different CH densities, that is, we can again see the reflection, the phase changes, the superposition effects of reflected and still incoming wave parts, and eventually a reflected wave including an enhanced as well as a depleted part, just with opposite phase and smaller amplitude values with regard to the incoming wave. The main difference to the situation in which the enhanced and the depleted amplitude are both equal to $1.1$ is first, the very small density values during the superposition process close to the CH, and second, the larger enhanced part of the reflected wave (resulting from small initial depletion amplitudes and the phase change).

In Figure \ref{1D_var_depl_widths} one can see the temporal evolution of a CW-CH interaction including constant initial enhancement and depletion amplitudes ($\rho_{IAE}=1.1$, $\rho_{IAD}=0.9$), a constant CH density ($\rho_{CH}=0.1$), a constant initial width for the enhanced part of the incoming wave ($w_{E}=0.1$) but different initial widths for the depleted part of the incoming wave ($w_D=0.1,0.15,0.2$). Again, we see a similar behaviour as in the parameter studies described above in Figures \ref{1D_var_CH_density} and \ref{1D_var_depl_ampl}. The main difference (as can be expected) can be seen in the width of the reflected enhanced part, that is, the larger the initial depleted width of the incoming wave, the larger the width and the amplitude of the enhanced reflected wave part. 

Figure \ref{1D_var_CHB_gradient} shows the temporal evolution of a CW-CH interaction with a constant initial density profile and varying CHB widths (sharp gradient and two smooth CHB gradients). The results show that a sharp gradient leads to a larger enhanced reflected wave part and a smaller width of the enhanced reflection, whereas a very smooth CHB gradient results in smaller enhanced reflected density amplitudes and a larger width of the enhanced reflection. 

\begin{figure}[ht!]
\centering\includegraphics[width=\linewidth]{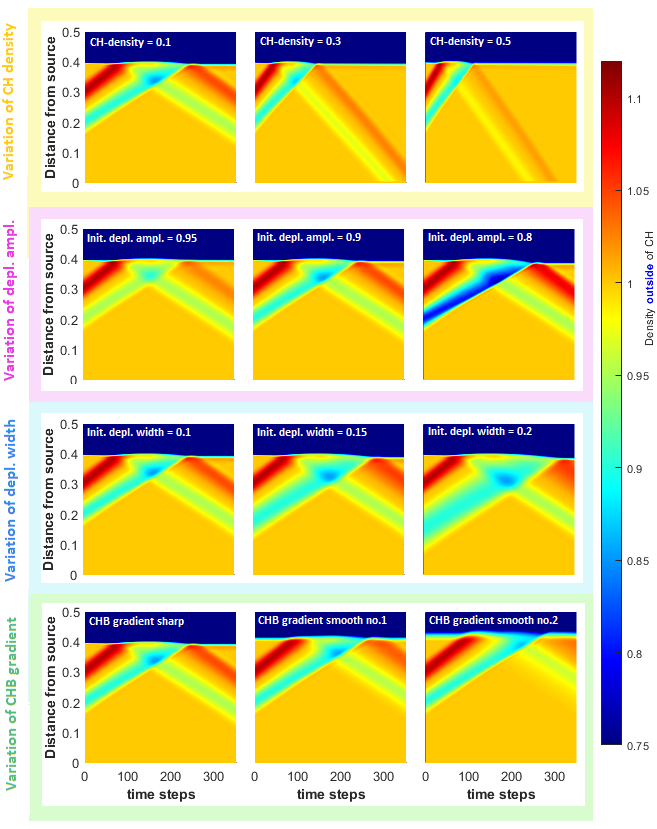}
\caption{Time-distance plots for different CH densities (first panel), different initial depletion amplitudes (second panel), different initial depletion widths (third panel), and different CHB widths (fourth panel).}
\label{1D_var_CH_density_timedist}
\end{figure}

Overall, one can see that the strongest interaction effects can be obtained by combining large initial amplitudes (limited to linear and weakly non-linear waves), small CH densities, large initial widths, and sharp CHB gradients. Figure \ref{1D_var_CH_density_timedist} shows these effects in the corresponding time-distance plots.

\section{Simulation results 2D}

\subsection{Initial conditions}

\begin{figure}[ht!]
\centering\includegraphics[width=\linewidth]{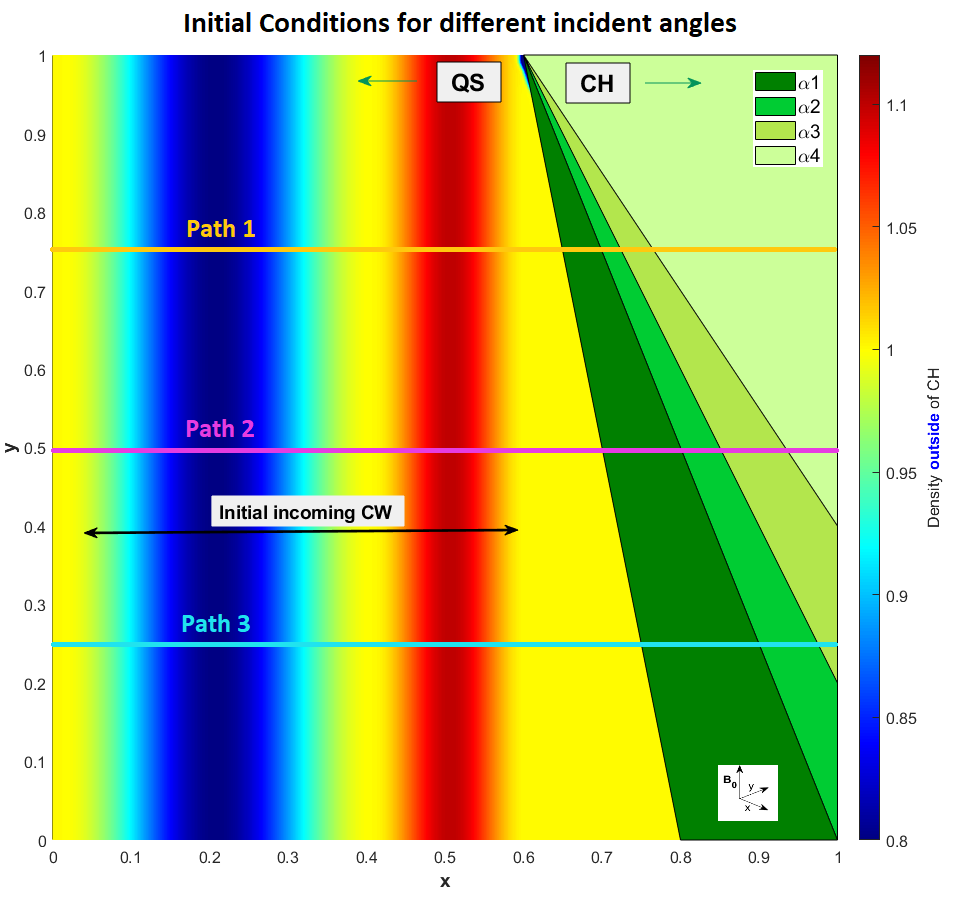}
\caption{Initial conditions for a 2D case of CW-CH interaction with a realistic density profile for the incoming wave ($\rho_{IAE}=1.1$, $\rho_{IAD}=0.8$, $w_E=0.1$, $w_D=0.2$), a constant CH density ($\rho_{CH}=0.1$), and four different incident angles ($\alpha_1=80^\circ$, $\alpha_2=70^\circ$, $\alpha_3=65^\circ$, and $\alpha_4=55^\circ$.). The corresponding time-distance plots for this interaction are generated along the depicted paths (Path1, Path2, and Path3).}
\label{2D_init_cond}
\end{figure}

Now that we have analysed the basic features in the 1D case of CW-CH interaction, it is crucial to study the 2D case with a realistic density profile (enhanced and depleted incoming wave parts) and different incident angles. As an initial setup for the incoming wave we choose those parameters that showed the strongest interaction effects in the 1D case, that is, $\rho_{IAE}=1.1$, $\rho_{IAD}=0.8$, $w_E=0.1$, $w_D=0.2$, and a sharp gradient representing the CHB. In addition to that, we choose three different paths along which the corresponding time-distance plots are created, in order to address the observational situations in which usually a clear and unique path for the propagation direction of the wave cannot be defined. In Figure \ref{2D_init_cond} one can see this initial 2D setup including the depleted part of the incoming wave (blue vertical structure), the enhanced part of the incoming wave (red vertical structure), the three paths along which we create the time-distance plots (Path 1, Path 2, and Path 3), and the four different incident angles ($\alpha_1$, $\alpha_2$, $\alpha_3$, and $\alpha_4$).

\subsection{Analysis of 2D simulation results and corresponding time-distance plots}

\begin{figure*}[ht!]
\centering\includegraphics[width=\textwidth]{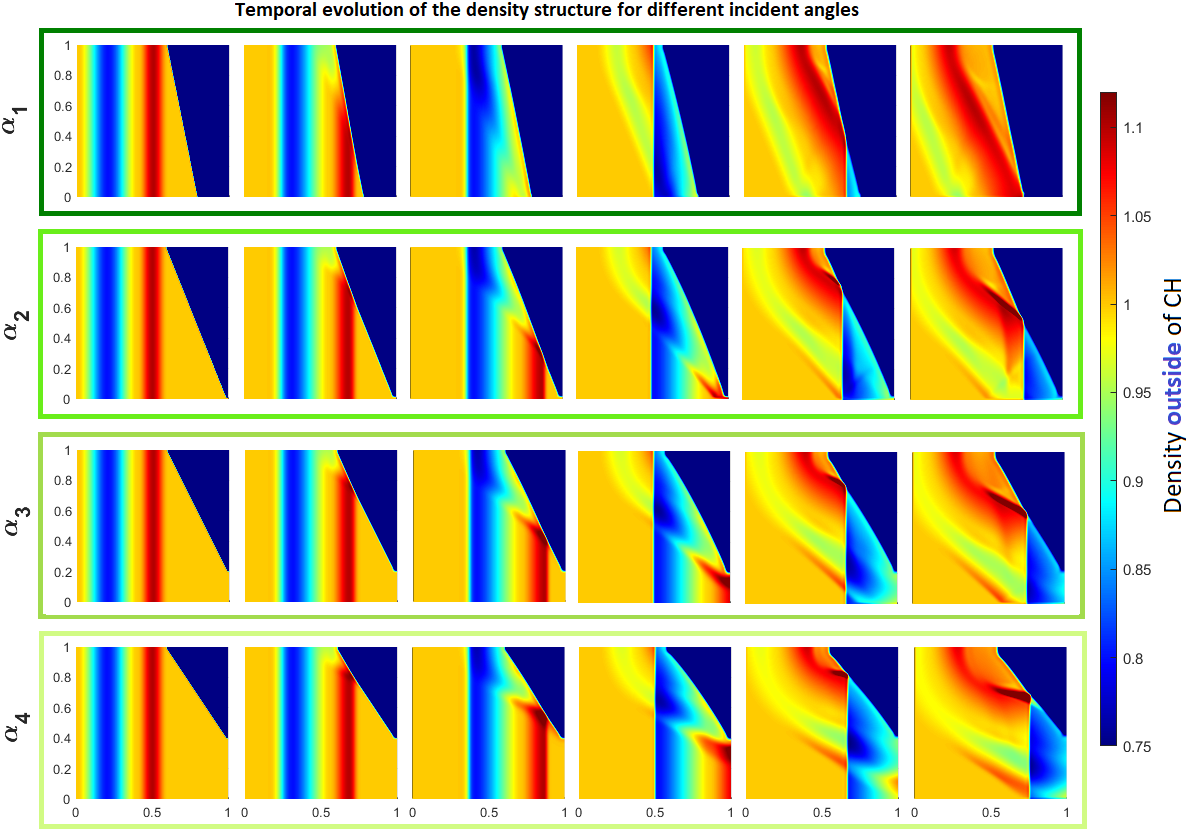}
\caption{Temporal evolution of the 2D density distribution for a CW-CH interaction with a realistic density profile that includes a depleted (blue vertical structure) and an enhanced wave part (red vertical structure), and four different incident angles, $\alpha_1=80^\circ$, $\alpha_2=70^\circ$, $\alpha_3=65^\circ$, and $\alpha_4=55^\circ$. (The temporal evolution shown in the last panel can also be seen in Movie 2.)}
\label{2D_var_diff_angles}
\end{figure*}

For the analysis of the simulation results in the 2D case, we focus on analysing the temporal evolution of the density structure and specifically on studying the corresponding time-distance plots as they are supposed to be used to directly compare the simulated interaction effects to actual observed interaction events. Figure \ref{2D_var_diff_angles} shows the temporal evolution of the 2D density distribution of a CW-CH interaction including a realistic density profile and four different incident angles. One can see that larger incident angles (such as $\alpha_1=80^\circ$ and $\alpha_2=70^\circ$) in combination with a realistic initial density profile lead to a more linear structured reflected area with reflection areas that hardly reach the large enhanced amplitudes of the incoming wave. Smaller incident angles (such as $\alpha_1=65^\circ$ and $\alpha_2=55^\circ$), on the other hand, result in enhanced reflected amplitudes that are much larger than the incoming ones and the reflected area exhibits a much more bended density structure. There is one more important implication that can be drawn from the temporal evolution of the density structure in Figure \ref{2D_var_diff_angles}. The strongly enhanced reflection (dark red structures at the end of the temporal density evolution) does neither move along the path of the incoming wave nor along the reflected wave path, but moves instead, due to superposition effects and the continuously incoming and interacting CW, in a somewhat perpendicular direction to the incoming wave while constantly changing its bended structure. This is an important information for the interpretation of observational data of CW-CH interactions effects.

\begin{figure*}[ht!]
\centering\includegraphics[width=0.99\textwidth]{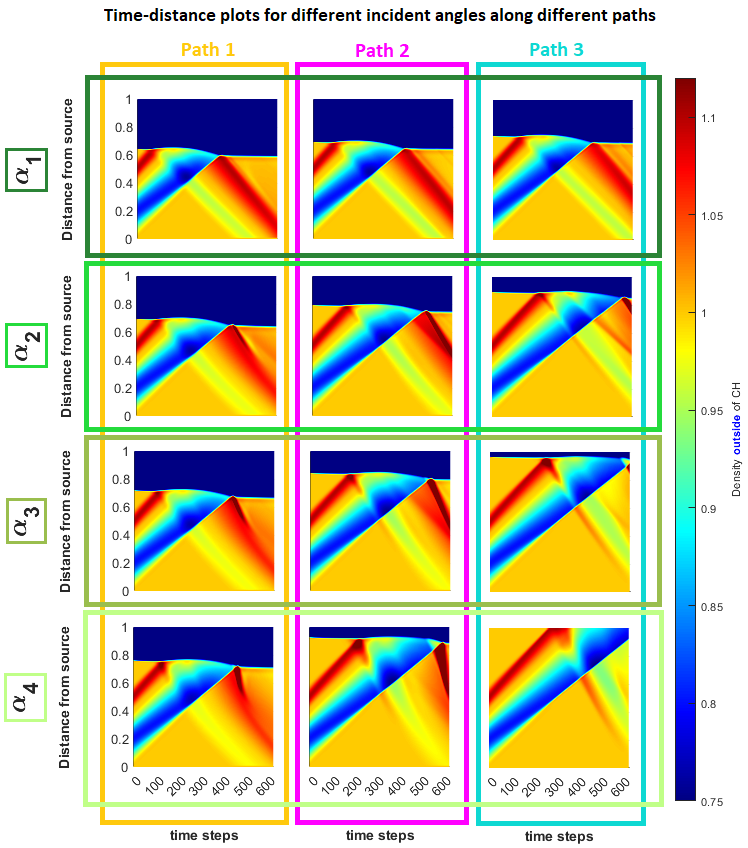}
\caption{Time-distance plots based on the temporal evolution of the density structure in Figure \ref{2D_var_diff_angles} along the three different paths (Path1, Path2, and Path3) described in the initial conditions of Figure \ref{2D_init_cond}. }
\label{2D_time_dist_var_angle}
\end{figure*}

\begin{figure*}
\centering\includegraphics[width=0.85\textwidth]{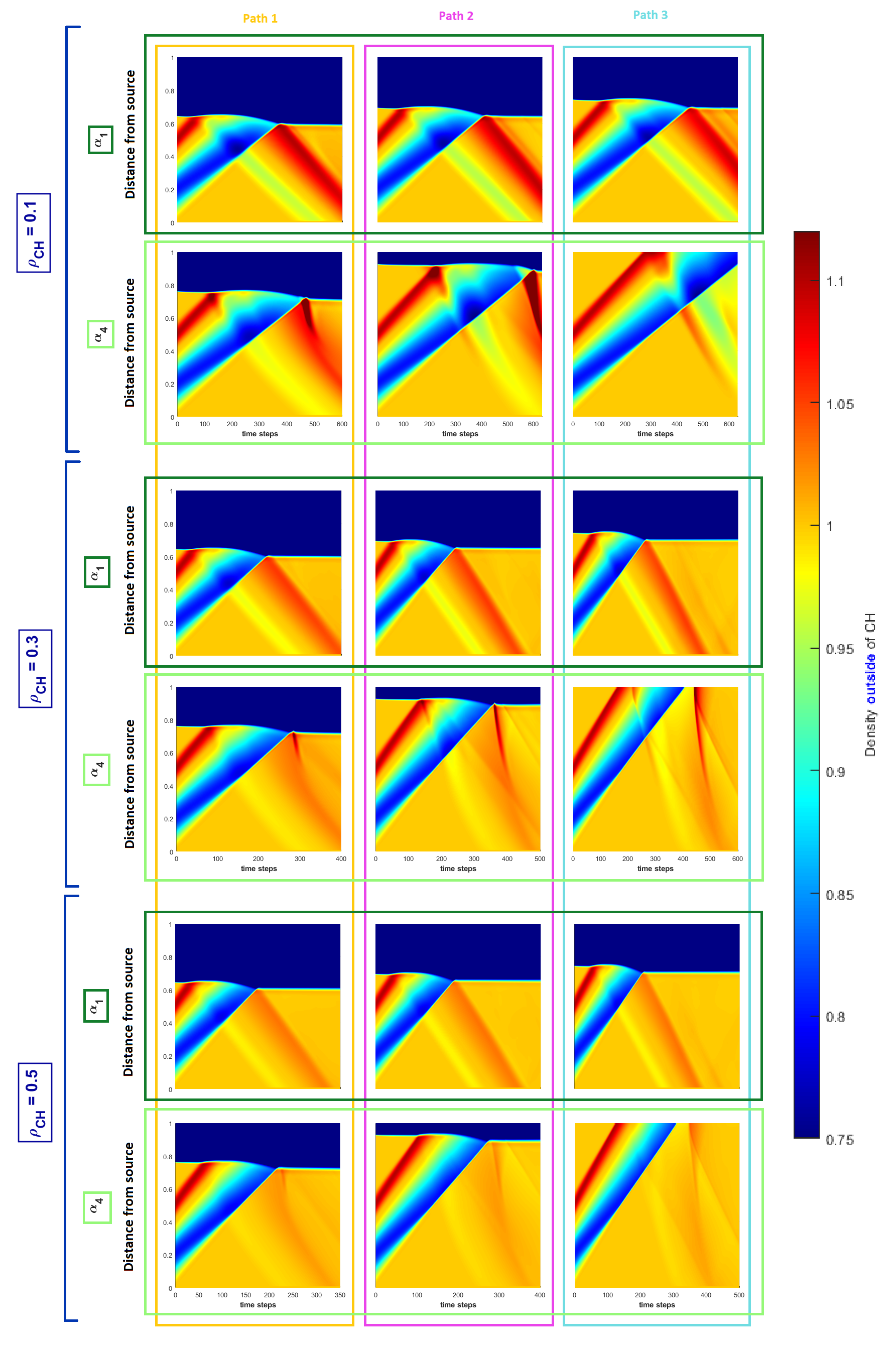}
\caption{Time-distance plots for the smallest and the largest incident angle, $\alpha_1$ and $\alpha_4$, along the three different paths (Path 1, Path 2, and Path 3), and for three different CH densities, $\rho_{CH}=0.1$, $\rho_{CH}=0.3$, and $\rho_{CH}=0.5$. The amplitudes of the initial density profile are $\rho_{IAE}=1.1$ and $\rho_{IAD}=0.8$, the initial widths are $w_E=0.1$, and $w_D=0.2$, and we used a constant CH density, $\rho_{CH}=0.1$.}
\label{2D_all_CH_extr_angles}
\end{figure*}

In Figure \ref{2D_time_dist_var_angle} one can see the time-distance plots based on the temporal evolution of the density structure in Figure \ref{2D_var_diff_angles} which were created along the three different paths (Path1, Path2, and Path3) described in the initial conditions of Figure \ref{2D_init_cond}. The first panel in Figure \ref{2D_time_dist_var_angle} shows the CW-CH interaction including the incident angle $\alpha_1=80^\circ$ and a CH density of $\rho_{CH}=0.1$ along the three different paths. As was already observed in Figure \ref{2D_var_diff_angles}, the shape of the reflected wave is almost as linear as the incoming one and the density amplitude of the reflected wave does not reach the values of the initial incoming wave. Also, the shape of the reflected wave stays more or less the same along the three different paths, whereas the enhanced amplitude decreases from Path1 to Path 3 for the situation including the angle $\alpha_1$. In the case of the incident angle $\alpha_2$, the situation already looks slightly different. One can see that the shape of the enhanced reflected wave part starts getting bent ($\alpha_2$ along Path1) and that this bent shape exhibits an even larger amplitude and a more complex structure along the other paths ($\alpha_2$ along Path2 and Path3). Overall, one can see that first, the smaller the incident angle, the larger the enhanced amplitude of the reflected wave and the more bent its density structure, and second, the farther the path is from the first interaction point (e.g., Path1 is close and Path3 is farther away), the more complex the reflected interaction structure and the larger the enhanced amplitudes.

A common method that is used to derive the phase speed of the reflected wave from observational data is to draw a line along the assumed propagation direction and generate a time-distance plot out of it. This method, however, is valid only in case the path for the incoming wave is exactly the same one as the slot chosen for the reflected wave. But in our case we have to consider what we see in the 2D density distribution of Figure \ref{2D_var_diff_angles}, namely the change of the shape of the reflected wave and the propagation direction which is definitely not the same as the one of the incoming wave. These results should also be considered when interpreting observational time-distance plots of CW-CH interaction.

In this study, we used the same path for the incoming and the reflected wave to create the time-distance plots, which is why we have to be careful regarding the interpretation of the actual reflected phase speed. But even if we try to determine a path along which the largely enhanced part of the reflected wave propagates, one also has to consider the bendend density structure and the continuous change of this structure during the interaction process. This means that one single slot for the whole propagation period of the reflected wave will not be sufficient to analyse the phase speed comprehensively in general, neither in the simulations nor in the observations.

What can be seen in all the time-distance plots of Figure \ref{2D_time_dist_var_angle} is the dark blue structure in the centre which depicts the period in which the reflected part of the enhanced incoming wave undergoes a phase change (from enhanced to depleted) and interacts with the depleted and still incoming wave. This superposition leads to a depleted density value that is smaller than the depleted amplitude of the initial incoming wave. During this period no enhanced reflected wave part is visible, however, this dark blue structure is not completely symmetric, it exhibits a small bump that points into the direction towards the upper left part of the plots. This bump, which continues into the direction of the green and yellow area, depicts the motion of the first reflected wave part that travels through the still incoming rear part of the wave until it reaches (due to superposition effects) the smallest density values and finally propagates as a slightly depleted reflection (light green structure) away from the CH. In the cases of the smaller incident angles, $\alpha_3$ and $\alpha_4$, one can also see a small dark red structure at the left side of the large depleted area, at around $t=100$ for Path 1 and at around $t=220$ for Path 2 (see also $t_3$ and $t_4$ in Figure \ref{2D_var_diff_angles}). This structure arises due to the small incident angle of the incoming wave (and therefore of the reflected wave) in combination with the still incoming enhanced wave part. In Figure \ref{2D_var_diff_angles} one can see that this dark red structure moves parallel to the incoming wave and explains therefore the superposition of enhanced wave part directly at the CHB.

Until now he have analysed time-distance plots with only a small CH density, that is, $\rho_{CH}=0.1$. In the last part of the section we therefore want to discuss the differences to situations including larger CH densities, that is, $\rho_{CH}=0.3$ and $\rho_{CH}=0.5$. From \citep{Piantschitsch2021} we know that the CH density determines the angles at which the phase of the reflected wave changes, therefore it is relevant to investigate the time-distance plots for different CH densities. Figure \ref{2D_all_CH_extr_angles} shows the time-distance plots for the smallest and the largest incident angle, $\alpha_1$ and $\alpha_4$, along the three different paths, Path 1, Path 2, and Path 3, and for three different CH densities, $\rho_{CH}=0.1$, $\rho_{CH}=0.3$, and $\rho_{CH}=0.5$. The first case in which $\rho_{CH}=0.1$ has already been discussed in Figure \ref{2D_time_dist_var_angle}, but in this plot we now compare it to the situations including larger CH densities. The time-distance plots for $\rho_{CH}=0.3$ and $\rho_{CH}=0.5$ show first and foremost that the interaction features, that is e.g., the large depleted area in the centre and the reflected wave amplitude, are not that strong as in the case of $\rho_{CH}=0.1$. In none of these cases the enhanced reflected amplitude reaches a value that is larger than the incoming enhanced amplitude, not even for small incident angles. However, a small CH density is not sufficient to achieve a large reflected amplitude, it has to be combined with a sufficiently small incident angle. 

Overall, and this is one of the key results of this paper, this means that an initial density profile that includes an enhanced as well as a depleted part, together with a sufficiently small incident angle and a small CH density lead to large reflected amplitudes and also to large phase speeds with respect to the incoming phase speed. If only one of these conditions is not satisfied a large reflected density amplitude cannot be reached. 

\subsection{Analytical expressions and comparison to simulation results in the 2D case}

As in the 1D case, we now want to compare the 2D simulation results to the analytical expressions derived in \citet{Piantschitsch2021} and make use again of the Equations (\ref{srinc})-(\ref{rho_refl_ratio}). As already mentioned above, these equations provide us information about the reflection coefficient which describes the changes of the amplitude of the reflected wave with respect to the incoming wave and depends only on the density contrast and the incident angle. By combining the Equations (\ref{srinc}) and (\ref{ffunct}) we can see that the reflection coefficient is real if $\theta_{\rm I}\ge  \cos^{-1} \left(\sqrt{\rho_{\rm c}}\right)$ and imaginary otherwise. For the comparison with the density values in the simulations we focus on situations including a real reflection coefficient. In the case of a small CH density of $\rho_{CH}=0.1$ this requirement is only met for the incident angle $\alpha_1$, for a larger CH density, such as $\rho_{CH}=0.5$, the reflection coefficient $R$ is real for all the incident angles used in this study. All these cases fulfill the condition that the incident angle is larger than $\cos^{-1} \left(\sqrt{\rho_{\rm c}}\right)$. For $\rho_{CH}=0.1$ and the incident angle $\alpha_4=55^\circ$ the reflection coefficient $R(55^\circ,0.1)\approx-0.46$. That implies that the minimum density value during the interaction, which is the sum of the incoming depleted amplitude and $R$ times the incoming enhanced density amplitude minus the background density, is approximately $0.754$ ( $\rho_{IAD} + R(\rho_{IAE}-\rho_0)\approx0.754$ ). This value is consistent with the density values of the depleted area in the centre of the time-distance plots in the first panel of Figure \ref{2D_time_dist_var_angle} and shows once more the agreement between the simulation results and the analytical expressions. An analogous comparison can be performed, for instance, for a large CH density such as $\rho_{CH}=0.5$ and all four incident angles $\alpha_1=80^\circ$, $\alpha_2=70^\circ$, $\alpha_3=65^\circ$, and $\alpha_4=55^\circ$. Here, we obtain the following values for the corresponding reflection coefficients: $R(80^\circ,0.5)\approx-0.1635$, $R(70^\circ,0.5)\approx-0.1369$, $R(65^\circ,0.5)\approx-0.1187$, and $R(55^\circ,0.5)\approx-0.0049$. If we use these analytically derived coefficients again to calculate the minimum depletion value, which occurs when the first part of the reflected wave interacts with the still incoming depletion, we obtain minimum density values of approximately $0.7837$, $0.7863$, $0.7881$, and $0.7995$ for $\alpha_1$, $\alpha_2$, $\alpha_3$, and $\alpha_4$. Again, these values show good agreement with the density values from the simulations as one can see in the last panel of Figure \ref{2D_all_CH_extr_angles}.

The angle that distinguishes the situation of a real reflection coefficient from an imaginary one in Equation (\ref{ffunct}) was already derived in \citet{Piantschitsch2021} and is called Critical angle, $\theta_C$,

\begin{equation}\label{thetc}
    \theta_{\rm C}={\rm cos^{-1}} \left(\sqrt{\rho_{\rm c}}\right).
\end{equation}

This angle separates the situation of full reflection from a situation that includes transmission, that is, incident angles above the Critical angle lead to transmitted waves whereas incident angles below the Critical angle result in pure reflections (for details see Figure 7 and Figure 8 in \citet{Piantschitsch2021}). This angle is properly defined when $\rho_{02}<\rho_{01}$ (which is the case for CHs) because the argument of $\rm cos^{-1}$ must always be between $-1$ and $1$. (Here, $\rho_{01}$ denotes the density in the first medium and $\rho_{02}$ denotes the density in the second medium.) In the case of $\rho_{01}<\rho_{02}$, there is no Critical angle, and this corresponds, for example, to any CW that propagates in the coronal medium and interacts with coronal loops which exhibit a higher density than their environment.

In \citet{Piantschitsch2021} analytical expressions for other specific angles were derived too. These angles show how the CH density determines whether an incoming wave gets totally reflected or partially transmitted, respectively, and at which angle the phase of the reflected wave changes (enhanced into depleted wave or vice versa). A crucial angle in this context is the so called Brewster angle, $\theta_B$,

\begin{equation}\label{thetb}
    \theta_{\rm B}={\rm cos^{-1}} \left(\sqrt{\frac{\rho_{\rm c}}{1+\rho_{\rm c}}}\right),
\end{equation}

where $\rho_{\rm c}$ denotes the density contrast (ratio of the CH density to the density of the quiet Sun). The Brewster angle represents a situation in which perfect transmission takes place, that is, there is no reflection at the CHB at all. Another important incident angle which separates enhanced from depleted reflection is that so-called Phase inversion angle, $\theta_P$,

\begin{equation}\label{thetS}
    \theta_{\rm P}={\rm cos^{-1}} \left(\sqrt{\frac{\rho_{\rm c}+\rho^2_{\rm c}}{1+\rho^2_{\rm c}}}\right),
\end{equation}

where $\rho_{\rm c}$ denotes again the density contrast. For linear waves this implies that an enhanced incoming wave with an incident angle between the Brewster angle and the Phase inversion angle gets reflected as an enhanced wave, whereas incoming waves that exhibit an incident angle outside of this range undergo a phase change and propagate as depleted reflections (for details see \citet{Piantschitsch2021}). For a CH density of $\rho_{CH}=0.1$, for instance, we know that the Brewster angle, $\theta_B$, is equal to $72.5^\circ$, the Phase inversion angle, $\theta_P$, is equal to $70.8^\circ$, and the Critical angle, $\theta_C$, is equal to $71.6^\circ$. One can see that in the case of a small CH density these values are very close to each other, also meaning that only a small change in the incident
angle is sufficient to turn a case of full transmission into
a case of no transmission at all. This is a crucial information for
the interpretation of CW–CH interaction effects in observational data as well.

These results are based on a linear approach and only purely enhanced incoming waves have been considered. However, one has to keep in mind that the phase change depending on the density contrast and the incident angle in combination with a realistic density profile have to be considered when interpreting the simulated interaction results in this study. Nonetheless, the analytical expressions for these angles provide important information about the behaviour of the reflected wave for different CH densities and show good agreement with the simulations performed in this study. 

Another interesting aspect in the course of studying CW-CH interaction, besides the properties of the reflected wave, is the transmission coefficient and how it is related to the reflection coefficient. Studying the properties of the transmitted wave in detail is out of the scope of this study, but we want to mention that in \citet{Piantschitsch2020} and \citet{Piantschitsch2021} we have already published theoretical results about the properties of the transmitted wave during a CW-CH interaction event and about how the transmission coefficient is related to the reflection coefficient in such situations.

\section{Conclusions}

In this study, we perform for the first time 2D MHD simulations of a CW-CH interaction including a realistic initial wave density profile that consists of an enhanced as well as a depleted wave part. We vary several initial parameters, such as the initial density amplitudes of the incoming wave, the CH density, and the CHB width, and we analyse the effects of different incident angles on the interaction features. The corresponding time-distance plots are used to detect specific features of the incoming and the reflected wave. 

The main results are summarised as follows:

    \begin{enumerate}
    
    \item For the first time, a density profile including an enhanced and a depleted part has been used to simulate the incoming wave in a CW-CH interaction process. The resulting interaction features are significantly different to those obtained by simulations using a purely enhanced pulse as initial incoming wave (see Figure \ref{compare_1D_single_pulse_realist_profile} and Movie 1). Among these differences are, e.g., a depletion area in the centre of the time-distance plots in which no wave part above background density level is visible and large reflected density amplitudes with respect to the incoming ones (see Figures \ref{2D_var_diff_angles} - \ref{2D_time_dist_var_angle} and Movie 2).
    \vspace{0.3cm}

    \item We performed parameters studies, varying the initial density amplitudes, the initial widths of the wave, the CH density, and the CHB width, and we found that small CH densities, large initial amplitudes, large initial widths and a steep CHB gradient lead to the strongest interaction results (see Figures \ref{1D_var_CH_density} - \ref{1D_var_CHB_gradient}).
    \vspace{0.3cm}

    \item By varying the incident angle in the 2D simulations we found that small incident angles lead to stronger superposition effects and therefore result in larger reflected density amplitudes and a more complex density distribution of the CW-CH interaction process (see Figures \ref{2D_var_diff_angles} and \ref{2D_time_dist_var_angle}, and Movie 2). 
    \vspace{0.3cm}

    \item As a key result of the simulations in this study, we found that a large reflected density amplitude, that is visible in some observational interaction events, can be obtained by a sufficiently small CH density in combination with a small incident angle and a realistic initial density profile of the incoming wave. If only one of these conditions is not satisfied a large reflected density amplitude cannot be reached (see Figures \ref{2D_time_dist_var_angle} and \ref{2D_all_CH_extr_angles}). This shows again the importance of including an enhanced as well as a depleted wave part in the initial setup of the incoming wave.
    \vspace{0.3cm}

     \item The density contrast plays a crucial role in two respects: First, a small CH density enhances the effects (amplitudes) for the reflected wave in a direct way (see Figure \ref{1D_var_CH_density}). Second, the density contrast also determines the angle which separates situations of full reflection from those which include transmission (see Equation (\ref{thetc})) and also defines the angle at which the wave undergoes a phase change (see Equation (\ref{thetS})) and therefore implicitly influences the properties of the reflected wave in an additional way.
    \vspace{0.3cm}

     \item The parameter studies in this study are supposed to be the initial step within a tool that reconstructs actual observational interaction events with the help of MHD simulations. By comparing the simulated time-distance plots to those created from observations, we aim in a further step to derive interaction parameters from the observed interaction events which usually cannot be obtained directly from the measurements.  
    \vspace{0.3cm}

    In this paper, we aim to combine information about observational measurements, such as the CH density, the initial density amplitudes of CWs, and CHB widths, with numerical MHD simulations of CW-CH interactions and theoretical results that provide analytical expressions for incident angles which influence the interaction process. We want to emphasize that in this study we are interested in linear and weakly non-linear CWs with a compression factor of around 1.1 or less. The interaction results naturally change for interactions including strongly non-linear waves that lead to the evolution of shocks and other features. 

    We also have to keep in mind, that we are considering an idealised situation including zero gas pressure and a homogeneous magnetic field. (A more realistic magnetic field topology will be included in future studies). Another idealisation in this paper is the simplified shape of the CH. The influence of different CH geometries on CW-CH interaction effects is out of the scope of this paper but will be addressed in a follow-up study.

    The parameter studies in this paper are considered to be a first step in reconstructing an actual interaction event. In a next step, we will compare the simulated time-distance plots to actual observations and try to derive interaction parameters from an observed CW-CH interaction event. By analysing the observational time-distance plots and comparing them to the time-distance plots based on the results of our numerical simulations we will be able to derive interaction parameters from the observed CW-CH interaction event that can usually not be determined directly.
  
    We believe that the results in this study can impact other simulations of coronal dynamics, such as the interaction of waves with prominences \citep[e.g.,][]{Liakh2023} or other processes in which the initial density profile of the CW plays a crucial role. 

    \end{enumerate}

%\begin{figure*}
%\includegraphics[width=0.99\textwidth]{Init_cond_circ_prop_resize.png}
%\caption{Time-distance plots for }
%\label{init_cond_circ}
%\end{figure*}

%\begin{figure*}
%\includegraphics[width=0.99\textwidth]{comparison_single_pulse_dimming.png}
%\caption{Time-distance plots for }
%\label{init_cond_circ}
%\end{figure*}

%\begin{figure*}
%\includegraphics[width=0.99\textwidth]{comparison_paths_time_distance_circle_resize.png}
%\caption{Time-distance plots for }
%\label{init_cond_circ}
%\end{figure*}

\begin{acknowledgements}

This research was funded by the Austrian Science Fund (FWF): Erwin-Schr\"odinger fellowship J4624-N. For the purpose of Open Access, the author has applied a CC BY public copyright licence to any Author Accepted Manuscript (AAM) version arising from this submission. This work was supported by the Austrian Science Fund (FWF): I3955-N27. JT and RS acknowledge support from  the R+D+i project PID2020-112791GB-I00, financed by MCIN/AEI/10.13039/501100011033. SGH acknowledges funding by the Austrian Science Fund (FWF): Erwin-Schr\"odinger fellowship J-4560.

\end{acknowledgements}

\bibliographystyle{aa}      % basic style, author-year citations
\bibliography{paper_reconst_event_cit}   % name your BibTeX data base

\begin{thebibliography}{48}
\expandafter\ifx\csname natexlab\endcsname\relax\def\natexlab#1{#1}\fi

\bibitem[{{Afanasyev} \& {Zhukov}(2018)}]{afanasyev2018}
{Afanasyev}, A.~N. \& {Zhukov}, A.~N. 2018, \aap, 614, A139

\bibitem[{{Chandra} {et~al.}(2022){Chandra}, {Chen}, {Devi}, {Joshi}, \&
  {Ni}}]{Chandra2022}
{Chandra}, R., {Chen}, P.~F., {Devi}, P., {Joshi}, R., \& {Ni}, Y.~W. 2022,
  Galaxies, 10, 58

\bibitem[{{Cranmer}(2009)}]{Cranmer2009}
{Cranmer}, S.~R. 2009, Living Reviews in Solar Physics, 6, 3

\bibitem[{{Del Zanna} \& {Bromage}(1999)}]{DelZanna1999}
{Del Zanna}, G. \& {Bromage}, B.~J.~I. 1999, \jgr, 104, 9753

\bibitem[{{Del Zanna} \& {Mason}(2018)}]{DelZanna2018}
{Del Zanna}, G. \& {Mason}, H.~E. 2018, Living Reviews in Solar Physics, 15, 5

\bibitem[{{Delaboudini{\`e}re} {et~al.}(1995){Delaboudini{\`e}re}, {Artzner},
  {Brunaud}, {Gabriel}, {Hochedez}, {Millier}, {Song}, {Au}, {Dere}, {Howard},
  {Kreplin}, {Michels}, {Moses}, {Defise}, {Jamar}, {Rochus}, {Chauvineau},
  {Marioge}, {Catura}, {Lemen}, {Shing}, {Stern}, {Gurman}, {Neupert},
  {Maucherat}, {Clette}, {Cugnon}, \& {van Dessel}}]{Delaboudiniere1995}
{Delaboudini{\`e}re}, J.~P., {Artzner}, G.~E., {Brunaud}, J., {et~al.} 1995,
  \solphys, 162, 291

\bibitem[{{Domingo} {et~al.}(1995){Domingo}, {Fleck}, \&
  {Poland}}]{Domingo1995}
{Domingo}, V., {Fleck}, B., \& {Poland}, A.~I. 1995, \ssr, 72, 81

\bibitem[{{Doschek} {et~al.}(1997){Doschek}, {Warren}, {Laming}, {Mariska},
  {Wilhelm}, {Lemaire}, {Sch{\"u}hle}, \& {Moran}}]{Doschek1997}
{Doschek}, G.~A., {Warren}, H.~P., {Laming}, J.~M., {et~al.} 1997, \apjl, 482,
  L109

\bibitem[{{Downs} {et~al.}(2021){Downs}, {Warmuth}, {Long}, {Bloomfield},
  {Kwon}, {Veronig}, {Vourlidas}, \& {Vr{\v{s}}nak}}]{Downs2021}
{Downs}, C., {Warmuth}, A., {Long}, D.~M., {et~al.} 2021, \apj, 911, 118

\bibitem[{{Gopalswamy} {et~al.}(2009){Gopalswamy}, {Yashiro}, {Temmer},
  {Davila}, {Thompson}, {Jones}, {McAteer}, {Wuelser}, {Freeland}, \&
  {Howard}}]{gopal2009}
{Gopalswamy}, N., {Yashiro}, S., {Temmer}, M., {et~al.} 2009, \apjl, 691, L123

\bibitem[{{Heinemann} {et~al.}(2021){Heinemann}, {Saqri}, {Veronig},
  {Hofmeister}, \& {Temmer}}]{Heinemann2021}
{Heinemann}, S.~G., {Saqri}, J., {Veronig}, A.~M., {Hofmeister}, S.~J., \&
  {Temmer}, M. 2021, \solphys, 296, 18

\bibitem[{{Heinemann} {et~al.}(2019){Heinemann}, {Temmer}, {Heinemann},
  {Dissauer}, {Samara}, {Jer{\v{c}}i{\'c}}, {Hofmeister}, \&
  {Veronig}}]{heinemann2019}
{Heinemann}, S.~G., {Temmer}, M., {Heinemann}, N., {et~al.} 2019, \solphys,
  294, 144

\bibitem[{{Heinemann} {et~al.}(2018){Heinemann}, {Temmer}, {Hofmeister},
  {Veronig}, \& {Vennerstr{\o}m}}]{Heinemann2018}
{Heinemann}, S.~G., {Temmer}, M., {Hofmeister}, S.~J., {Veronig}, A.~M., \&
  {Vennerstr{\o}m}, S. 2018, \apj, 861, 151

\bibitem[{{Hofmeister} {et~al.}(2022){Hofmeister}, {Asvestari}, {Guo},
  {Heidrich-Meisner}, {Heinemann}, {Magdalenic}, {Poedts}, {Samara}, {Temmer},
  {Vennerstrom}, {Veronig}, {Vr{\v{s}}nak}, \&
  {Wimmer-Schweingruber}}]{Hofmeister2022}
{Hofmeister}, S.~J., {Asvestari}, E., {Guo}, J., {et~al.} 2022, \aap, 659, A190

\bibitem[{{Hofmeister} {et~al.}(2018){Hofmeister}, {Veronig}, {Temmer},
  {Vennerstrom}, {Heber}, \& {Vr{\v{s}}nak}}]{Hofmeister2018}
{Hofmeister}, S.~J., {Veronig}, A., {Temmer}, M., {et~al.} 2018, Journal of
  Geophysical Research (Space Physics), 123, 1738

\bibitem[{{Hofmeister} {et~al.}(2020){Hofmeister}, {Veronig}, {Poedts},
  {Samara}, \& {Magdalenic}}]{Hofmeister2020}
{Hofmeister}, S.~J., {Veronig}, A.~M., {Poedts}, S., {Samara}, E., \&
  {Magdalenic}, J. 2020, \apjl, 897, L17

\bibitem[{{Kienreich} {et~al.}(2013){Kienreich}, {Muhr}, {Veronig},
  {Berghmans}, {De Groof}, {Temmer}, {Vr{\v{s}}nak}, \&
  {Seaton}}]{kienreichetal2013}
{Kienreich}, I.~W., {Muhr}, N., {Veronig}, A.~M., {et~al.} 2013, \solphys, 286,
  201

\bibitem[{{Krieger} {et~al.}(1973){Krieger}, {Timothy}, \&
  {Roelof}}]{Krieger1973}
{Krieger}, A.~S., {Timothy}, A.~F., \& {Roelof}, E.~C. 1973, \solphys, 29, 505

\bibitem[{{Liakh} {et~al.}(2023){Liakh}, {Luna}, \& {Khomenko}}]{Liakh2023}
{Liakh}, V., {Luna}, M., \& {Khomenko}, E. 2023, \aap, 673, A154

\bibitem[{{Liu} {et~al.}(2019){Liu}, {Wang}, {Lee}, \& {Shen}}]{liu19}
{Liu}, R., {Wang}, Y., {Lee}, J., \& {Shen}, C. 2019, \apj, 870, 15

\bibitem[{{Liu} \& {Ofman}(2014)}]{Wei2014}
{Liu}, W. \& {Ofman}, L. 2014, \solphys, 289, 3233

\bibitem[{{Long} {et~al.}(2008){Long}, {Gallagher}, {McAteer}, \&
  {Bloomfield}}]{Long2008}
{Long}, D.~M., {Gallagher}, P.~T., {McAteer}, R.~T.~J., \& {Bloomfield}, D.~S.
  2008, \apjl, 680, L81

\bibitem[{{Mancuso} {et~al.}(2021){Mancuso}, {Bemporad}, {Frassati},
  {Barghini}, {Giordano}, {Telloni}, \& {Taricco}}]{Mancuso2021}
{Mancuso}, S., {Bemporad}, A., {Frassati}, F., {et~al.} 2021, \aap, 651, L14

\bibitem[{{Moreton} \& {Ramsey}(1960)}]{Moreton60}
{Moreton}, G.~E. \& {Ramsey}, H.~E. 1960, \pasp, 72, 357

\bibitem[{{Muhr} {et~al.}(2011){Muhr}, {Veronig}, {Kienreich}, {Temmer}, \&
  {Vr{\v{s}}nak}}]{Muhr2011}
{Muhr}, N., {Veronig}, A.~M., {Kienreich}, I.~W., {Temmer}, M., \&
  {Vr{\v{s}}nak}, B. 2011, \apj, 739, 89

\bibitem[{{Nolte} {et~al.}(1976){Nolte}, {Krieger}, {Timothy}, {Gold},
  {Roelof}, {Vaiana}, {Lazarus}, {Sullivan}, \& {McIntosh}}]{Nolte1976}
{Nolte}, J.~T., {Krieger}, A.~S., {Timothy}, A.~F., {et~al.} 1976, \solphys,
  46, 303

\bibitem[{{Olmedo} {et~al.}(2012){Olmedo}, {Vourlidas}, {Zhang}, \&
  {Cheng}}]{olmedoetal2012}
{Olmedo}, O., {Vourlidas}, A., {Zhang}, J., \& {Cheng}, X. 2012, \apj, 756, 143

\bibitem[{{Piantschitsch} \& {Terradas}(2021)}]{Piantschitsch2021}
{Piantschitsch}, I. \& {Terradas}, J. 2021, \aap, 651, A67

\bibitem[{{Piantschitsch} {et~al.}(2020){Piantschitsch}, {Terradas}, \&
  {Temmer}}]{Piantschitsch2020}
{Piantschitsch}, I., {Terradas}, J., \& {Temmer}, M. 2020, \aap, 641, A21

\bibitem[{{Piantschitsch} {et~al.}(2018{\natexlab{a}}){Piantschitsch},
  {Vr{\v{s}}nak}, {Hanslmeier}, {Lemmerer}, {Veronig}, {Hernandez-Perez}, \&
  {{\v{C}}alogovi{\'c}}}]{Piantschitsch2018a}
{Piantschitsch}, I., {Vr{\v{s}}nak}, B., {Hanslmeier}, A., {et~al.}
  2018{\natexlab{a}}, \apj, 857, 130

\bibitem[{{Piantschitsch} {et~al.}(2018{\natexlab{b}}){Piantschitsch},
  {Vr{\v{s}}nak}, {Hanslmeier}, {Lemmerer}, {Veronig}, {Hernandez-Perez}, \&
  {{\v{C}}alogovi{\'c}}}]{Piantschitsch2018b}
{Piantschitsch}, I., {Vr{\v{s}}nak}, B., {Hanslmeier}, A., {et~al.}
  2018{\natexlab{b}}, \apj, 860, 24

\bibitem[{{Piantschitsch} {et~al.}(2017){Piantschitsch}, {Vr{\v{s}}nak},
  {Hanslmeier}, {Lemmerer}, {Veronig}, {Hernandez-Perez},
  {{\v{C}}alogovi{\'c}}, \& {{\v{Z}}ic}}]{Piantschitsch2017}
{Piantschitsch}, I., {Vr{\v{s}}nak}, B., {Hanslmeier}, A., {et~al.} 2017, \apj,
  850, 88

\bibitem[{{Podladchikova} {et~al.}(2019){Podladchikova}, {Veronig},
  {Podladchikova}, {Dissauer}, {Vr{\v{s}}nak}, {Saqri}, {Piantschitsch}, \&
  {Temmer}}]{podladchikova2019}
{Podladchikova}, T., {Veronig}, A.~M., {Podladchikova}, O., {et~al.} 2019, in
  EGU General Assembly Conference Abstracts, EGU General Assembly Conference
  Abstracts, 9793

\bibitem[{{Riley} {et~al.}(2015){Riley}, {Linker}, \& {Arge}}]{riley2015}
{Riley}, P., {Linker}, J.~A., \& {Arge}, C.~N. 2015, Space Weather, 13, 154

\bibitem[{{Samara} {et~al.}(2022){Samara}, {Magdaleni{\'c}}, {Rodriguez},
  {Heinemann}, {Georgoulis}, {Hofmeister}, \& {Poedts}}]{Samara2022}
{Samara}, E., {Magdaleni{\'c}}, J., {Rodriguez}, L., {et~al.} 2022, \aap, 662,
  A68

\bibitem[{{Saqri} {et~al.}(2020){Saqri}, {Veronig}, {Heinemann}, {Hofmeister},
  {Temmer}, {Dissauer}, \& {Su}}]{Saqri2020}
{Saqri}, J., {Veronig}, A.~M., {Heinemann}, S.~G., {et~al.} 2020, \solphys,
  295, 6

\bibitem[{{Thompson} {et~al.}(1998){Thompson}, {Plunkett}, {Gurman}, {Newmark},
  {St. Cyr}, \& {Michels}}]{thompson1998}
{Thompson}, B.~J., {Plunkett}, S.~P., {Gurman}, J.~B., {et~al.} 1998, \grl, 25,
  2465

\bibitem[{{T{\'o}th} \& {Odstr{\v{c}}il}(1996)}]{toth1996}
{T{\'o}th}, G. \& {Odstr{\v{c}}il}, D. 1996, Journal of Computational Physics,
  128, 82

\bibitem[{van Leer(1979)}]{VanLeer1979}
van Leer, B. 1979, Journal of Computational Physics, 32, 101

\bibitem[{van Leer(1984)}]{VanLeer1984}
van Leer, B. 1984, SIAM Journal on Scientific and Statistical Computing, 5, 1

\bibitem[{{Veronig} {et~al.}(2010){Veronig}, {Muhr}, {Kienreich}, {Temmer}, \&
  {Vr{\v{s}}nak}}]{veronig2010}
{Veronig}, A.~M., {Muhr}, N., {Kienreich}, I.~W., {Temmer}, M., \&
  {Vr{\v{s}}nak}, B. 2010, \apjl, 716, L57

\bibitem[{{Veronig} {et~al.}(2006){Veronig}, {Temmer}, {Vr{\v{s}}nak}, \&
  {Thalmann}}]{Veronig2006}
{Veronig}, A.~M., {Temmer}, M., {Vr{\v{s}}nak}, B., \& {Thalmann}, J.~K. 2006,
  \apj, 647, 1466

\bibitem[{{Vr{\v{s}}nak} \& {Luli{\'c}}(2000)}]{vrsnaklulic2000}
{Vr{\v{s}}nak}, B. \& {Luli{\'c}}, S. 2000, \solphys, 196, 157

\bibitem[{{Wang}(2000)}]{wang2000}
{Wang}, Y.~M. 2000, \apjl, 543, L89

\bibitem[{{Warmuth}(2015)}]{Warmuth2015}
{Warmuth}, A. 2015, Living Reviews in Solar Physics, 12, 3

\bibitem[{{Warmuth} {et~al.}(2004){Warmuth}, {Vr{\v{s}}nak}, {Magdaleni{\'c}},
  {Hanslmeier}, \& {Otruba}}]{Warmuth2004}
{Warmuth}, A., {Vr{\v{s}}nak}, B., {Magdaleni{\'c}}, J., {Hanslmeier}, A., \&
  {Otruba}, W. 2004, \aap, 418, 1101

\bibitem[{{Wu} {et~al.}(2001){Wu}, {Zheng}, {Wang}, {Thompson}, {Plunkett},
  {Zhao}, \& {Dryer}}]{wu2001}
{Wu}, S.~T., {Zheng}, H., {Wang}, S., {et~al.} 2001, \jgr, 106, 25089

\bibitem[{{Zhou} {et~al.}(2022){Zhou}, {Shen}, {Tang}, {Zhou}, {Duan}, \&
  {Tan}}]{Zhou2022}
{Zhou}, X., {Shen}, Y., {Tang}, Z., {et~al.} 2022, \aap, 659, A164

\end{thebibliography}

%\begin{figure*}
%\includegraphics[width=\textwidth]%{extreme_case_1D_2D_3D_resize.png}
%\caption{Time-distance plots for }
%\label{init_cond_circ}
%\end{figure*}

%\begin{figure*}
%\centering \includegraphics[width=0.55\textwidth]%{sketch_angles_update3.png}
%\caption{\small Schematic representation of the interaction %between an oblique incoming wave and a CHB, depending on the %angle of the incident wave, $\theta_{\rm I}$. Five different %scenarios are possible (assuming that the incoming wave is an %enhancement): a) no transmission in the $x$-direction and the %reflection is a depletion, b) no transmission and the %reflection is an enhancement, c) transmission in the $x$-%direction and the reflection is an enhancement, d) perfect %transmission, also meaning that there is no reflection at all, %e) transmission in the $x$-direction and the reflection is a %depletion. }\label{angle_comp1}
%\end{figure*}

\end{document}